\documentclass[aps,pre,twocolumn]{revtex4}
\usepackage{graphics}
\usepackage{graphicx}

\begin{document}


\title{Dynamical chaos in nonlinear Schr\"odinger models with subquadratic power nonlinearity}


\author{Alexander~V.~Milovanov${}^{1,3}$ and Alexander~Iomin${}^{2,3}$}

\affiliation{${}^1$ENEA National Laboratory, Centro~Ricerche~Frascati, I-00044 Frascati, Rome, Italy}
\affiliation{${}^2$Department of Physics, Technion$-$Israel Institute of Technology, 32000 Haifa, Israel}
\affiliation{${}^3$Max Planck Institute for the Physics of Complex Systems, D-01187 Dresden, Germany}




\begin{abstract} 
We devise an analytical method to deal with a class of nonlinear Schr\"odinger lattices with random potential and subquadratic power nonlinearity. An iteration algorithm is proposed based on multinomial theorem, using Diophantine equations and a mapping procedure onto a Cayley graph. Based on this algorithm, we were able to obtain several hard results pertaining to asymptotic spreading of the nonlinear field beyond a perturbation theory approach. In particular, we show that the spreading process is subdiffusive and has complex microscopic organization involving both long-time trapping phenomena on finite clusters and long-distance jumps along the lattice consistent with L\'evy flights. The origin of the flights is associated with the occurrence of degenerate states in the system; the latter are found to be a characteristic of the subquadratic model. The limit of quadratic power nonlinearity is also discussed and shown to result in a delocalization border, above which the field can spread to long distances on a stochastic process and below which it is Anderson localized similarly to a linear field. 
\end{abstract}


\maketitle

\section{Introduction} 

There has been a stream of publications initiated by Shepelyansky \cite{Sh93}, suggesting that a weak nonlinearity might destroy Anderson localization of dynamical chaos in disordered systems, if the strength of nonlinear interaction exceeds a certain threshold strength. The phenomenon was analyzed theoretically and demonstrated numerically on the basis of the nonlinear Schr\"odinger (i.e., Gross-Pitaevskii) equation (NLSE) with random potential \cite{PS,Wang,Flach,Skokos,Fishman,Iomin,EPL}. The key issue here is the structure of the nonlinear term, which defines the interactions among the components of the wave field. 
Most previous theoretical studies have assumed that the nonlinear frequency shift resulting from these interactions is directly proportional to intensity of the wave field and that the relationship between nonlinear frequency and intensity is local in that the frequency at a point is defined by intensity at the {\it same} point. A justification for this has referred to Taylor expansion of frequency over the amplitude of the nonlinear field, leading to a quadratic correction in the first order. Then in a mean-field approximation one writes the NLSE as ({\it e.g.}, Refs. \cite{Gross,Sulem}) 
\begin{equation}
i\hbar\frac{\partial\psi_n}{\partial t} = \hat{H}_L\psi_n + \beta |\psi_n|^{2} \psi_n,
\label{QNLSE} 
\end{equation}
where $\hat{H}_L$ is the Hamiltonian of the linear problem; $\psi _n = \psi (n, t)$ is the complex wave function, which may vary with time, $t$; $n = 1,\dots, N$ is the wave number; and $\beta$ characterizes the strength of nonlinearity. On the other hand, the localization-delocalization phenomena discussed in, {\it e.g.}, Refs. \cite{DF2017,Horn2017,Ashourvan2017,PRE21,Garbet21} have shown that in some systems the relationship between nonlinear frequency and intensity could be {\it nonlocal} involving a distribution of wave processes across an extended area in wave number space. A common feature to those systems is the existence of a long-range, nonlocal ordering competing with the order parameter $\propto |\psi_n|^2$. Writing the nonlinear frequency shift resulting from nonlocal ordering as 
\begin{equation}
\Delta\omega_{\rm NL} = \beta^\prime \sum_{|n - n^\prime| = 1}^{N}\chi ({n - n^\prime)}{|\psi_{n^\prime}|^{2}},
\label{NLFS} 
\end{equation}
one is led to an NLSE with distributed nonlinearity
\begin{equation}
i\hbar\frac{\partial\psi_n}{\partial t} = \hat{H}_L\psi_n + \beta^\prime \Big(\sum_{|n - n^\prime| = 1}^{N}\chi ({n - n^\prime)}{|\psi_{n^\prime}|^{2}}\Big)\psi_n,
\label{DSNL} 
\end{equation}
where $\beta^\prime$ is a coefficient, which absorbs the impact of nonlinear interactions in the presence of nonlocal order. 
In the above, $\chi = \chi (n-n^\prime)$ is a response function, which, for statistically homogeneous, isotropic systems, depends solely on $|n-n^\prime|$. Equation~(\ref{DSNL}) is a starting point to explore a rich variety of localization-delocalization phenomena in nonlinear systems with distributed interactions. 
Among the possible examples we mention systems of strongly coupled transport barriers in L-mode tokamak plasma, the so-called plasma staircase \cite{DF2017,Horn2017,Ashourvan2017,PRE21,Garbet21,DF2010,DF2015}. Other examples include gravitational optics \cite{Segev,Iomin21}, cosmological dark waves \cite{EPL22+}, nonlinear fracton dynamics \cite{Naka,UFN}, arrays of coupled Josephson junctions \cite{Martinoli,PRB02}, coherent excitation transport in biological macromolecules and charge transport in organic semiconductors \cite{Mingalev96,Dyre}, just to address some. Note that we directly associate nonlocality with the nonlinear term [via the convolution in Eq.~(\ref{NLFS})]. In this respect, NLSE~(\ref{DSNL}) is different from the analogue equations considered in, {\it e.g.}, Refs. \cite{Iomin21,Weitzner,PLA2005,Korabel2007,Zonca2006}, where nonlocality is introduced into the linear (dispersion) term, leading, in those cases, to a fractional modification of NLSE. We do not introduce those modifications here.  

In this paper's work, we are interested in a reduction of NLSE~(\ref{DSNL}) to a nonlinear Schr\"odinger equation with subquadtratic power nonlinearity. We argue that the ensuing NLSE [see Eq.~(\ref{1}) below] could be an efficient and powerful tool to characterize systems of coupled nonlinear oscillators with long-range dependence, being, at the same time, a valuable alternative to known fractional ventures (Refs. \cite{Weitzner,PLA2005,Korabel2007}; Refs. \cite{UFN,Chapter} for reviews). Mathematically, the use of subquadratic power might, however, be very nontrivial, leading, in some cases, to considerable technical difficulties \cite{PRE21,PRE19}. It is, therefore, the goal of the present study to sort out the existing mathematical issues, and to establish a general analytical framework to deal with this type of dynamical equations beyond a perturbation theory approach. In particular, we propose an iteration procedure, making it possible to expand the subquadratic nonlinearity into a multinomial power series, and to analyze this series analytically by representing it as an infinite sequence of the Cayley graphs with properly chosen coordination numbers. In this fashion, we were able to obtain several hard results concerning the existence (or nonexistence) of asymptotic transport processes pertaining to the nonlinear field. A brief account of some of these investigations has been reported previously \cite{EPL22}.   

The paper is organized as follows. The reduction of the distributed nonlinearity in Eq.~(\ref{DSNL}) to a subquadratic power nonlinearity is taken through steps in Sec.~II. The exact dynamical model pertaining to our work is defined in Sec.~III. The multinomial expansion of subquadratic nonlinearity is considered in Sec.~IV. In Sec.~V, we show how the partial nonlinearities arising at the different orders of this expansion could be represented graphically using a mapping procedure onto a Cayley tree. A scaling theory of field spreading is proposed in Sec.~VI, followed by a probabilistic picture of asymptotic transport, which is drawn in Sec.~VII with the aid of a bifractional diffusion equation. In Sec.~VIII, we discuss the limiting case of quadratic power nonlinearity, which is shown to be special in that it leads to both the strong and weak (percolation-style) chaos regimes depending on strength of nonlinear interaction. We summarize our findings in Sec.~IX.
   
\section{From distributed to subquadratic nonlinearity}

We consider a partial form of Eq.~(\ref{DSNL}), such that the response function is given by the power law $\chi (n-n^\prime) \propto 1/|n-n^\prime|^s$, where $s$ is a power exponent lying within $0 < s < 1$. This form of $\chi (n-n^\prime)$ can be motivated by a system of coupled nonlinear oscillators being close to a marginally stable ({\it i.e.}, critical) state, where dynamical fluctuations are supposed to be scale free \cite{Sornette2004,Pruessner,Asch2013}. The occurrence of such states in L-mode tokamak plasma has been a matter of comprehensive investigation ({\it e.g.}, Refs. \cite{Rhodes,Politzer,Newman,Sharma}). Then instead of Eq.~(\ref{DSNL}) one gets
\begin{equation}
i\hbar\frac{\partial\psi_n}{\partial t} = \hat{H}_L\psi_n + \beta_s \Big(\sum_{|n - n^\prime| = 1}^{N}\frac{|\psi_{n^\prime}|^{2}}{|n - n^\prime|^s}\Big) \psi_n,
\label{DSPL} 
\end{equation}  
where $\beta_s$ is a parameter characterizing the interaction problem, and the sum in round brackets is a discretization of the so-called Riemann-Liouville fractional integral \cite{Podlubny,Samko} of order $1-s$. Equation~(\ref{DSPL}) is combined with the normalization condition 
\begin{equation}
\sum_{n^\prime = 1}^N |\psi_{n^\prime}|^2 = 1,
\label{Norm} 
\end{equation}  
which represents the conservation of the total probability in the system. 

If the nonlinear field is spread across a large number of states $1 \ll \Delta n \leq N$, then the summation in Eq.~(\ref{Norm}) may be performed in the limits from $1$ to $\Delta n$ (because the remaining terms corresponding to $n^\prime$ running from $\Delta n$ to $N$ will be zero), leading to 
\begin{equation}
\sum_{n^\prime = 1}^{\Delta n} |\psi_{n^\prime}|^2 = 1.
\label{Norm2} 
\end{equation}  
If, however, the nonlinear field is such that it only slightly varies between the occupied states, then from Eq.~(\ref{Norm2}) one infers that the density of the probability is small and behaves with $\Delta n$ as $|\psi_{n^\prime}|^2 \sim |\psi_n|^2 \sim 1/\Delta n$. Furthermore, if $0 < s < 1$ and $|n - n^\prime| \gg 1$, then the power-law function $\chi (n-n^\prime) \propto 1/|n-n^\prime|^s$ may be thought of as a ``slow" function of the number of states. Taking this function out of the summation in Eq.~(\ref{DSPL}) and summing over $n^\prime$ from $1$ to $\Delta n$, in the view of Eq.~(\ref{Norm}) one gets 
\begin{equation}
i\hbar\frac{\partial\psi_n}{\partial t} = \hat{H}_L\psi_n + \beta_s (\Delta n)^{-s}\psi_n,
\label{DSPL+} 
\end{equation}  
where the nonlinear correction to frequency depends on the width of field distribution. 

\section{The model}

Eliminating $\Delta n$ with the aid of $|\psi_n|^2 \sim (\Delta n)^{-1}$, from Eq.~(\ref{DSPL+}) one is led to an NLSE of the form
\begin{equation}
i\hbar\frac{\partial\psi_n}{\partial t} = \hat{H}_L\psi_n + \beta_s |\psi_n|^{2s} \psi_n.
\label{1} 
\end{equation}
Equation~(\ref{1}) is the central equation of the present study. This equation suggests that the self-organization into a critical state \cite{Sornette2004,Asch2013} is a global process, though one which can be described by a local model with subquadratic power nonlinearity. The latter type of nonlinearity results in the nonlinear frequency shift 
\begin{equation}
\Delta\omega_{\rm NL} = \beta_s |\psi_n|^{2s}.
\label{NLSH} 
\end{equation}
For $s=1$, the paradigmatic model in Eq.~(\ref{QNLSE}) is recovered, with $\beta_s = \beta$. Concerning the Hamiltonian of the linear problem, $\hat{H}_L$, our main assumption is that $\hat{H}_L$ possesses a full basis of mutually orthogonal eigenfunctions, which we shall denote as $\phi_{n,k}$. By their definition, the eigenfunctions $\phi_{n,k}$ are supposed to obey $\hat H_{L} \phi_{n,k} = \omega_k \phi_{n,k}$, where $\omega_k$ denote the respective eigenfrequencies, and $k = 0,\pm 1,\pm 2,\dots$ is an integer counter. Orthogonality means that the following condition holds:
\begin{equation}
\sum _n \phi^*_{n,m}\phi_{n,k} = \delta_{m,k},
\label{Kro} 
\end{equation}
where $\delta_{m,k}$ is Kronecker's delta, and the star $^*$ denotes complex conjugate. For the sake of concreteness, we take $\hat{H}_L$ to be the familiar Anderson Hamiltonian in the tight-binding approximation \cite{And}, {\it i.e.},
\begin{equation}
\hat{H}_L\psi_n = \varepsilon_n\psi_n + V (\psi_{n+1} -2\psi_n + \psi_{n-1}),
\label{2} 
\end{equation}
for which the existence of a full basis of mutually orthogonal eigenstates is a well-established property \cite{And+,Kramer}. In the above, $V$ is hopping matrix element, $\psi_{n+1} - 2\psi_n + \psi_{n-1}$ mimics the dispersion term for next-neighbor jumps along the coordinate $n$, and $\varepsilon_n$ is the energy at site $n$. It is assumed that $\varepsilon_n$ is a random quantity and as such incorporates the spatial disorder of the lattice on which the interactions take place. Without loss in generality, we define that the shifted energy values $\varepsilon_n^\prime = \varepsilon_n - 2V$ are randomly distributed with zero mean across a finite energy interval $-W/2 \leq \varepsilon_n^\prime \leq W/2$, thus corresponding to the settings of Refs. \cite{PS,Fishman,EPL}. In the absence of randomness, NLSE~(\ref{1}) is fully integrable. For $\beta_s = 0$, all eigenstates pertaining to Eq.~(\ref{1}) with the Hamiltonian~(\ref{2}) are exponentially localized with the localization length $\lambda \approx 96 (V/W)^2$ \cite{Kramer}. In what follows, $\beta_s \geq 0$ and $V = \hbar = 1$ for simplicity.     

\section{Multinomial expansion}

Expanding $\psi_n$ over the basis functions, one writes
\begin{equation}
\psi_n = \sum_m \sigma_m (t) \phi_{n,m},
\label{3} 
\end{equation}
where $\sigma_m (t)$ are complex functions, which may depend on time $t$, and $m = 0,\pm 1,\pm 2,\dots$ is an integer counter. If $s=1$, then based on NLSE~(\ref{QNLSE}) one obtains a set of dynamical equations for $\sigma_m (t)$, valid for all $t \geq 0$. For this, one needs to substitute Eq.~(\ref{3}) into~(\ref{QNLSE}), multiply the both sides by $\phi^*_{n,k}$, and sum over $n$, using the orthonormality condition in Eq.~(\ref{Kro}). The result is 
\begin{equation}
i\dot{\sigma}_k - \omega_k \sigma_k = \beta \sum_{m_1, m_2, m_3} V_{k, m_1, m_2, m_3} \sigma_{m_1} \sigma^*_{m_2} \sigma_{m_3},
\label{4} 
\end{equation}
where 
the coefficients $V_{k, m_1, m_2, m_3}$ are given by
\begin{equation}
V_{k, m_1, m_2, m_3} = \sum_{n} \phi^*_{n,k}\phi_{n,m_1}\phi^*_{n,m_2}\phi_{n,m_3}
\label{5} 
\end{equation}
and characterize couplings among the nonlinear wave processes pertaining to Eq.~(\ref{4}). 
In the above we have used $\beta$ instead of $\beta_s$ consistently with the model in Eq.~(\ref{QNLSE}). 
Equation~(\ref{4}) describes a chain of coupled nonlinear oscillators with the Hamiltonian 
\begin{equation}
\hat H = \hat H_{0} + \hat H_{\rm int},
\label{Ham} 
\end{equation}
where 
\begin{equation}
\hat H_0 = \sum_k \omega_k \sigma^*_k \sigma_k
\label{6-0} 
\end{equation}
is the Hamiltonian of non-interacting harmonic oscillators, and 
\begin{equation}
\hat H_{\rm int} = \frac{\beta}{2} \sum_{k, m_1, m_2, m_3} V_{k, m_1, m_2, m_3} \sigma^*_k \sigma_{m_1} \sigma^*_{m_2} \sigma_{m_3}
\label{6+} 
\end{equation}
is the interaction Hamiltonian. Remark that we include self-interactions into $\hat H_{\rm int}$. It is understood that each nonlinear oscillator with the Hamiltonian   
\begin{equation}
\hat h_{k} = \omega_k \sigma^*_k \sigma_k + \frac{\beta}{2} V_{k, k, k, k} \sigma^*_k \sigma_{k} \sigma^*_{k} \sigma_{k}
\label{6+h} 
\end{equation}
and the equation of motion 
\begin{equation}
i\dot{\sigma}_k - \omega_k \sigma_k - \beta V_{k, k, k, k} \sigma_{k} \sigma^*_{k} \sigma_{k} = 0
\label{eq} 
\end{equation}
represents one nonlinear eigenstate in the system~(\ref{4}), identified by its wave number $k$, unperturbed frequency $\omega_k$ and the nonlinear frequency shift $\Delta \omega_{k} = \beta V_{k, k, k, k} \sigma_{k} \sigma^*_{k}$. 

If $s < 1$, then one needs to assess
\begin{equation}
|\psi_n|^{2s} = (\psi_n\psi_n^{*})^s = \left[\sum_{m_1,m_2} \sigma_{m_1} \sigma^{*}_{m_2} \phi_{n,m_1}\phi^*_{n,m_2}\right]^s,
\label{2s} 
\end{equation}
which represents a mathematical problem of general significance. Note, in this respect, that the algebraic form in Eq.~(\ref{2s}) behaves as a $C$-number functional form and adheres to the usual $C^*$-algebra (see, {\it e.g.}, Ref. \cite{Arveson} and the comment in Ref. \cite{Comm1}). 

If the exponent $s$ was a positive integer number$-$a type of nonlinearity considered in Refs. \cite{FL,DNC}$-$then the right-hand side of Eq.~(\ref{2s}) would be given by exact multinomial expansion \cite{Stegun}
\begin{equation}
|\psi_n|^{2s} = \sum_{\sum {q_{m_1,m_2}} = s}\mathcal{C}_s^{\dots q_{m_1,m_2}}\prod_{m_1, m_2} [\xi_{m_1,m_2}]^{q_{m_1,m_2}},
\label{Multinom} 
\end{equation}
where 
\begin{equation}
\mathcal{C}_s^{\dots q_{m_1,m_2}} = \frac{s!}{\prod_{m_1, m_2} [q_{m_1,m_2}!]}
\label{Coeff} 
\end{equation}
are multinomial coefficients, the sign $!$ stands for the factorial function, and we have denoted for simplicity
\begin{equation}
\xi_{m_1,m_2} = \sigma_{m_1} \sigma^*_{m_2} \phi_{n,m_1}\phi^*_{n,m_2}.
\label{Simp} 
\end{equation}
The summation in Eq.~(\ref{Multinom}) is performed over all combinations of natural numbers $q_{m_1,m_2}$ such that for each combination the sum $\sum_{m_1,m_2}{q_{m_1,m_2}}$ is $s$, {\it i.e.},  
\begin{equation}
\sum_{m_1,m_2}{q_{m_1,m_2}} = s.
\label{Sums} 
\end{equation}
For fractional values of $s$, the multinomial expansion in Eq.~(\ref{Multinom}) is not defined, yet one may devise \cite{PRE21,PRE19} its analytic continuation to $0 < s < 1$ by extending the factorial function to Euler's gamma function using $q_{m_1,m_2}! = \Gamma (q_{m_1,m_2}+1)$ and simultaneously relaxing the condition that the exponents $q_{m_1,m_2}$ must be integer. This analytic continuation to the interval $0 < s < 1$ is obtained {\it iteratively} over the increasing number of fractional exponents satisfying Eq.~(\ref{Sums}). That is, one starts with an approximation (call it first-order), when there is one and only one fractional exponent to be assumed in Eq.~(\ref{Sums}); followed by a more complex setting when the fractional exponents are just two and only two ({\it i.e.}, second-order approximation); and so forth. The goal is to assess the respective impacts of the nonlinearities arising at the various orders on the localization properties of the nonlinear field. As is shown below, this iteration procedure converges exponentially fast, making it possible to solve the localization problem in Eq.~(\ref{1}) {\it exactly} in all orders. We proceed as follows.  

\subsection{First-order approximation}

So in the first order Eq.~(\ref{Sums}) is satisfied, if {\it the} fractional exponent we are looking for [which, in this case, is the {\it only} fractional exponent contributing to Eq.~(\ref{Sums}), since the others are integer] is equal to $s$ exactly (because the sum of the remaining integer-valued exponents cannot add up to a fractional value). Then from Eq.~(\ref{Sums}) one infers that the sum of the integer-valued exponents is zero, and this is an exact result. Assume it is the exponent $q_{i,j}$ which takes the fractional value, {\it i.e.}, $q_{m_1,m_2} = s < 1$ for some $m_1 = i$ and $m_2 = j$. With this setting, we may establish
\begin{equation}
\sum_{m_1\ne i,m_2\ne j}{q_{m_1,m_2}} = 0.
\label{Dioph} 
\end{equation}
Equation~(\ref{Dioph}) is a Diophantine equation, which is a polynomial equation for which only integer solutions are sought. Noting that the exponents $q_{m_1,m_2}$ cannot take negative values [since otherwise the expanded field diverges for $|\psi_n|^2 \rightarrow +0$], the only way to satisfy~(\ref{Dioph}) is by setting all $q_{m_1,m_2}$ to zero, if $m_1 \ne i$, $m_2\ne j$; while keeping $q_{i,j}=s$ for $m_1 =i$, $m_2 = j$.
One sees that the product $\prod_{m_1, m_2}$ in Eq.~(\ref{Multinom}) is composed of one factor only, enabling 
\begin{equation}
\prod_{m_1, m_2} [\xi_{m_1,m_2}]^{q_{m_1,m_2}} \rightarrow [\xi_{m_1 = i,m_2 = j}]^{s}.
\label{Pro1} 
\end{equation}
It is understood that the polynomial in Eq.~(\ref{Multinom}) is homogeneous in that the sum of the exponents at each term is always $s$, as Eq.~(\ref{Sums}) shows. On the other hand, the property of homogeneity implies that {\it any} term pertaining to Eq.~(\ref{Multinom}) is in some sense representative of the whole. That means that there is no particular reason for which to prefer the very specific setting $m_1 = i, m_2 = j$ against other equivalent settings when choosing the fractional-valued $q_{m_1,m_2}$. The net result is that the condition $q_{i,j} = s$ can be satisfied in a countable number of ways within the range of variation of the parameters $m_1$ and $m_2$. Clearly, all such combinations would equally contribute to the series expansion in Eq.~(\ref{Multinom}). Then to account for these contributions one has to sum over the indexes $m_1$ and $m_2$. To this end, Eq.~(\ref{Multinom}) becomes 
\begin{equation}
|\psi_n|_{(1)}^{2s} = \sum_{{m_1,m_2}} [\xi_{m_1,m_2}]^{s},
\label{Homo} 
\end{equation}
where the subscript ``$_{(1)}$" reminds that the nonlinearity in Eq.~(\ref{Homo}) applies at the first order of multinomial expansion of $|\psi_n|^{2s}$. Also in writing Eq.~(\ref{Homo}) we have used   
\begin{equation}
\mathcal{C}_s^{\dots q_{i,j}} = \frac{\Gamma (s+1)}{\Gamma (q_{i,j} + 1)} = \frac{\Gamma (s+1)}{\Gamma (s + 1)} = 1.
\label{First} 
\end{equation}
Eliminating $\xi_{m_1,m_2}$ with the aid of Eq.~(\ref{Simp}), from the series expansion in Eq.~(\ref{Homo}) one arrives at 
\begin{equation}
|\psi_n|_{(1)}^{2s} = \sum_{{m_1,m_2}} \sigma^{s}_{m_1} \sigma_{m_2}^{*s} \phi^{s}_{n,m_1}\phi^{*s}_{n,m_2}.
\label{HF} 
\end{equation}

Applying the nonlinearity in Eq.~(\ref{HF}), it is straightforward to obtain dynamical equations for $\sigma_k (t)$ in the first order. For this, one needs to substitute (\ref{HF}) into NLSE~(\ref{1}), multiply the both sides of ensuing wave equation by $\phi^*_{n,k}$, and sum over $n$, using the orthonormality condition in Eq.~(\ref{Kro}). The result is  
\begin{equation}
i\dot{\sigma}_k - \omega_k \sigma_k = \beta_s \sum_{m_1, m_2, m_3} V_{k, m_1, m_2, m_3} \sigma^{s}_{m_1} \sigma_{m_2}^{* s} \sigma_{m_3},
\label{4s+} 
\end{equation}
where, similarly to Eq.~(\ref{5}), 
\begin{equation}
V_{k, m_1, m_2, m_3} = \sum_{n} \phi^*_{n,k}\phi^{s}_{n,m_1}\phi_{n,m_2}^{* s}\phi_{n,m_3}.
\label{5s+} 
\end{equation}
The interaction Hamiltonian corresponding to~(\ref{4s+}) is given by  
\begin{equation}
\hat H_{\rm int} = \frac{\beta_s}{1+s} \sum_{k, m_1, m_2, m_3} V_{k, m_1, m_2, m_3} \sigma^*_k \sigma_{m_1}^s \sigma^{*s}_{m_2} \sigma_{m_3}
\label{6s+} 
\end{equation}
and generalizes the Hamiltonian in Eq.~(\ref{6+}). 

\subsection{Second and higher orders}

Turning to the second-order approximation, one assumes that the number of fractional exponents pertaining to Eq.~(\ref{Sums}) is just two (and only two), whereas any other exponents (if there are such exponents) are given by integer values. All these exponents (whatever fractional or not) must, moreover, be nonnegative to ensure good behavior in the infrared limit, where the wave field vanishes. Because $s < 1$, one sees that the only possibility is that the sum of the fractional-valued exponents is exactly $s$, while the sum of the integer-valued exponents is zero. Denoting the fractional exponents as $q_{i_1,j_1}$ and $q_{i_2,j_2}$, one finds that Eq.~(\ref{Sums}) is split into two separate equations, that is, $q_{i_1,j_1} + q_{i_2,j_2} = s$ and the Diophantine equation 
\begin{equation}
\sum_{m_1\ne i_1, i_2}\sum_{m_2\ne j_1, j_2}{q_{m_1,m_2}} = 0,
\label{Dioph2} 
\end{equation}
from which it is deduced that all integer-valued exponents must be equal to zero, {\it i.e.}, $q_{m_1,m_2} = 0$ for $m_1\ne i_1, i_2$ and $m_2\ne j_1, j_2$. That means that the {\it only} two exponents that would meaningfully contribute to $\prod_{m_1, m_2}$ in Eq.~(\ref{Multinom}) are the fractional ones. By assumption these exponents are, precisely, $q_{i_1,j_1}$ and $q_{i_2,j_2}$. One sees that the product $\prod_{m_1, m_2}$ reduces to a product among two terms only, identified by the powers $q_{i_1,j_1}$ and $q_{i_2,j_2}$. One gets 
\begin{equation}
\prod_{m_1, m_2} [\xi_{m_1,m_2}]^{q_{m_1,m_2}} \rightarrow [\xi_{i_1,j_1}]^{q_{i_1,j_1}} [\xi_{i_2,j_2}]^{q_{i_2,j_2}}.
\label{Pro2} 
\end{equation}
This corresponds to a nonlinearity of the form 
\begin{equation}
|\psi_n|_{(2)}^{2s} = \sum_{{i_1,j_1}} \sum_{{i_2,j_2}} \Theta_{n,i_1,j_1,i_2,j_2}\sigma^{q_{i_1,j_1}}_{i_1} \sigma_{j_1}^{*q_{i_1,j_1}} \sigma^{q_{i_2,j_2}}_{i_2} \sigma_{j_2}^{*q_{i_2,j_2}},
\label{HF+2} 
\end{equation}
where, for simplicity,
\begin{equation}
\Theta_{n,i_1,j_1,i_2,j_2} = \theta_{i_1,j_1,i_2,j_2} \phi^{q_{i_1,j_1}}_{n,i_1}\phi^{*q_{i_1,j_1}}_{n,j_1} \phi^{q_{i_2,j_2}}_{n,i_2}\phi^{*q_{i_2,j_2}}_{n,j_2},
\label{HF+2T} 
\end{equation}
the subscript ``$_{(2)}$" signifies that the double summation in Eq.~(\ref{HF+2}) pertains to the second order of multinomial expansion, the coefficients $\theta_{i_1,j_1,i_2,j_2}$ are given by 
\begin{equation}
\theta_{i_1,j_1,i_2,j_2} = {\Gamma (s+1)}/[{\Gamma (q_{i_1,j_1} + 1)\Gamma (q_{i_2,j_2} + 1)]}
\label{HF+2TT} 
\end{equation}
and correspond to the multinomial coefficients in Eq.~(\ref{Coeff}), and $q_{i_2,j_2} = s - q_{i_1,j_1}$ consistently with Eq.~(\ref{Sums}). The nonlinearity in Eq.~(\ref{HF+2}) amends the first-order nonlinearity in Eq.~(\ref{HF}) to yield, in the second order, 
\begin{equation}
|\psi_n|^{2s} \simeq |\psi_n|_{(1)}^{2s} + |\psi_n|_{(2)}^{2s}.
\label{Sum2} 
\end{equation}

To obtain dynamical equations for $\sigma_k (t)$ in the second order, one needs to substitute~(\ref{Sum2}) into NLSE~(\ref{1}) and expand $|\psi_n|_{(1)}^{2s}$ and $|\psi_n|_{(2)}^{2s}$ using, respectively, Eqs.~(\ref{HF}) and~(\ref{HF+2}). 
Next one has to multiply the both sides of Eq.~(\ref{1}) by $\phi^*_{n,k}$ and sum over $n$, remembering that the basis functions $\phi_{n,m}$ are orthogonal. The result of this calculation is summarized by
\begin{equation}
i\dot{\sigma}_k - \omega_k \sigma_k = \beta_s\mathcal{{S}}_{k,(1)} + \beta_s\mathcal{{S}}_{k,(2)},
\label{Res2} 
\end{equation}
where
\begin{equation}
\mathcal{{S}}_{k,(1)} = \sum_{m_1, m_2, m_3} V_{k, m_1, m_2, m_3} \sigma^{s}_{m_1} \sigma_{m_2}^{* s} \sigma_{m_3}
\label{4s++} 
\end{equation}
and reproduces the triple sum $\sum_{m_1, m_2, m_3}$ on the right-hand side of Eq.~(\ref{4s+}); whereas $\mathcal{{S}}_{k,(2)}$ stands for the quintic polynomial 
\begin{equation}
\sum_{{i_1,j_1,i_2,j_2,m_3}} V_{k,i_1,j_1,i_2,j_2,m_3}\sigma^{q_{i_1,j_1}}_{i_1} \sigma_{j_1}^{*q_{i_1,j_1}} \sigma^{q_{i_2,j_2}}_{i_2} \sigma_{j_2}^{*q_{i_2,j_2}}\sigma_{m_3}
\label{Delta} 
\end{equation}
and absorbs corrections to $\mathcal{{S}}_{k,(1)}$ in the second order. In the above, $V_{k,i_1,j_1,i_2,j_2,m_3}$ are coefficients of the polynomial and are defined through
\begin{equation}
V_{k,i_1,j_1,i_2,j_2,m_3} = \sum_{n} \phi^*_{n,k} \Theta_{n,i_1,j_1,i_2,j_2} \phi_{n,m_3}.
\label{VK} 
\end{equation}

The iteration procedure devised above in the first and second orders may be extended to arbitrary order $\ell\geq 1$ of the multinomial expansion of $|\psi_n|^{2s}$. At each order $\ell$ one assumes that the number of fractional-valued exponents is equal to $\ell$ exactly, then it is found by solving a Diophantine equation that any eventual integer exponents must be equal to zero ensuring good behavior in the infrared limit. 
For each $\ell = 1,2,\dots$, one builds an interaction process characterized by the nonlinearity $|\psi_n|_{(\ell)}^{2s}$, which incorporates couplings among $2l\geq 2$ different waves (2 waves in the first order; 4 waves in the second order; etc.) If we adopt, for the reasons of formal ordering, that the coupling probability between two waves is characterized by a small parameter $\epsilon \ll 1$, then the interaction process corresponding to $|\psi_n|_{(1)}^{2s}$ has the order $\epsilon^1$, the interaction process corresponding to $|\psi_n|_{(2)}^{2s}$ has the order $\epsilon^2$, and so forth. One sees that the iteration algorithm deriving from the approximation 
\begin{equation}
|\psi_n|^{2s} \simeq \sum_{r = 1}^{\ell} |\psi_n|_{(r)}^{2s} 
\label{Approx} 
\end{equation}
converges exponentially fast with increasing $r = 1,2\dots$, making it possible to assess the asymptotic dynamics pertaining to NLSE~(\ref{1}) by allowing $\ell\rightarrow+\infty$ in Eq.~(\ref{Approx}).
    
\section{Mapping on a Cayley tree}

To demonstrate the existence (or nonexistence) of asymptotic transport at each finite order $\ell$, we employ a procedure already devised in Refs. \cite{EPL,PRE14} for $\ell = 1$, according to which one needs to investigate the fine structure of the interaction term in the corresponding dynamical equations for $\sigma_k (t)$. This is achieved by mapping the interaction term onto a Cayley tree \cite{Schroeder} with a suitably chosen coordination number, $z$. 

For $s=1$, the procedure is as follows \cite{EPL}. Each nonlinear oscillator with the Hamiltonian~(\ref{6+h}) and equation of motion~(\ref{eq}) is represented by a node on a Cayley graph; each such node is then connected to other nodes by a system of ``conducting" (able to transmit a wave) bonds. We distinguish between ingoing (corresponding to the amplitudes $\sigma^*_{m_i}$) and outgoing (corresponding to $\sigma_{m_i}$) bonds. An examination of dynamical Eq.~(\ref{4}) shows that there will be exactly three such bonds at each node, {\it i.e.}, one ingoing corresponding to the wave amplitude $\sigma^*_{m_2}$, and two outgoing bonds corresponding to the amplitudes $\sigma_{m_1}$ and $\sigma_{m_3}$. This generates an element of a Cayley tree with the coordination number $z=3$ (see Fig.~1, left). 

\begin{figure}
\includegraphics[width=0.51\textwidth]{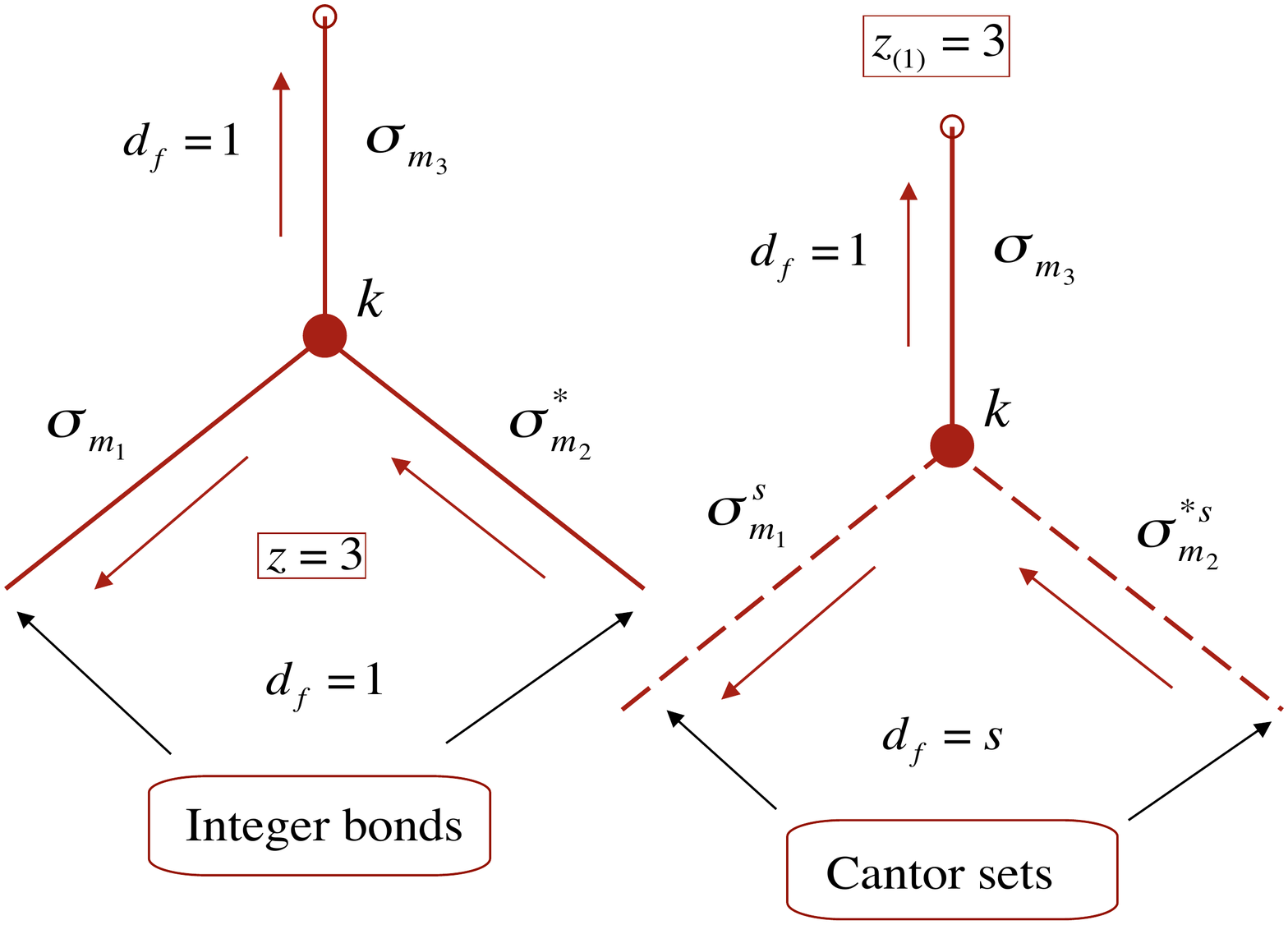}
\caption{\label{} Mapping dynamical equations~(\ref{4}) (left) and~(\ref{4s+}) (right) on a Cayley tree with the coordination number $z=3$. 
}
\end{figure}
\begin{figure}
\includegraphics[width=0.51\textwidth]{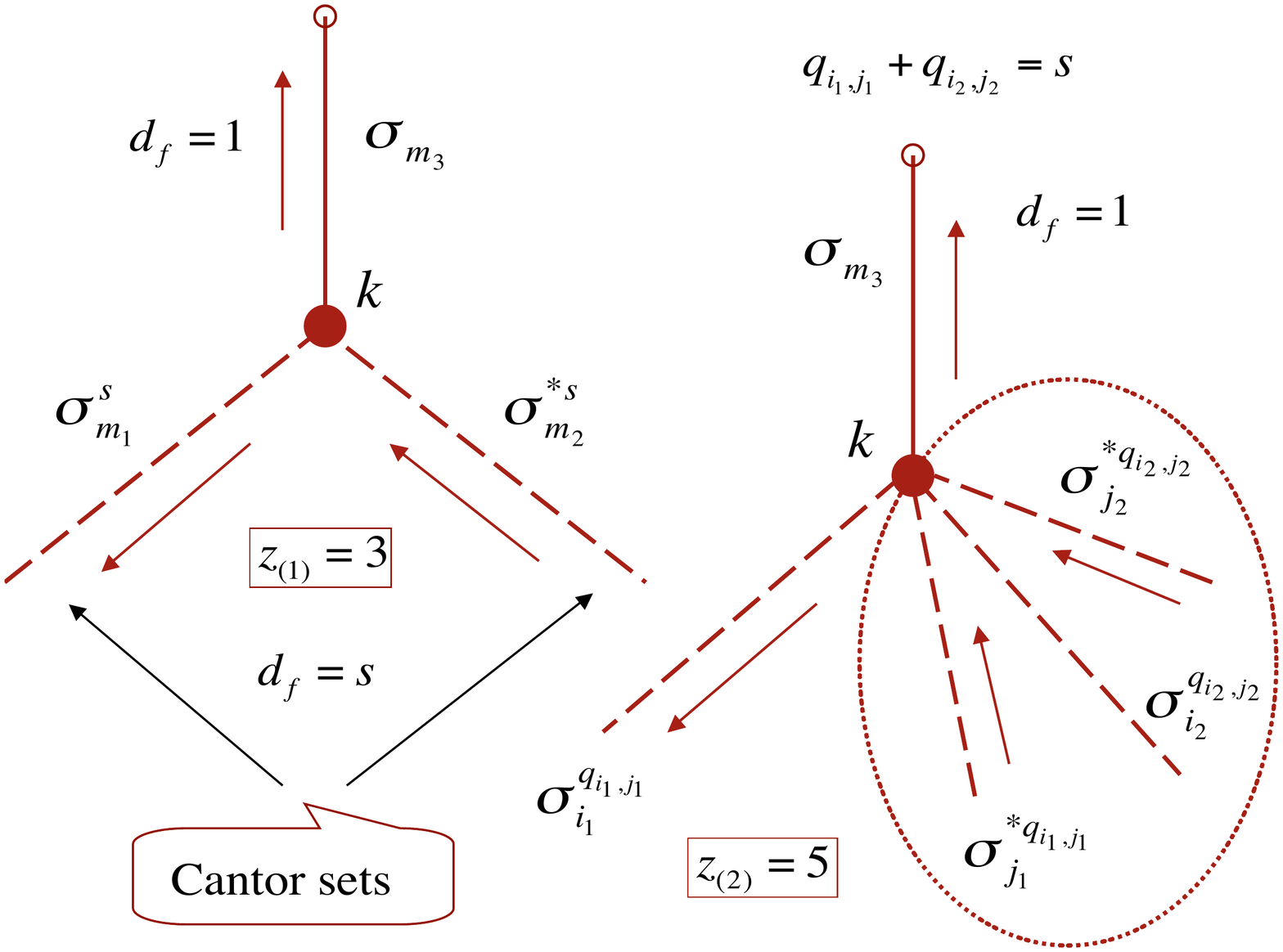}
\caption{\label{} Mapping dynamical equations for $\sigma_k (t)$ onto a Cayley tree in the first (left) and second (right) orders assuming $s < 1$. Dotted oval on the right-hand side encircles a degenerate state. 
}
\end{figure}

\subsection{The issue of disconnected bonds for $s < 1$}

If $s < 1$, then a similar mapping procedure could be applied to dynamical Eq.~(\ref{4s+}), though with that new feature \cite{PRE14} that the amplitudes $\sigma^{s}_{m_1}$ and $\sigma_{m_2}^{* s}$ are represented by everywhere disconnected bonds (Cantor sets), with the Hausdorff \cite{Feder} fractal dimension $d_f = s$. That means that in the first order of multinomial expansion one again obtains a Cayley tree with the coordination number $z_{(1)}=3$; yet, there is a particularity stemming from $s < 1$, namely, that two (and only two, out of three possible) bonds at each node are disconnected (see Fig.~1, right). 

In a classical approach, a bond being disconnected would imply it cannot transmit a wave, the result being that the spreading process is disrupted at the Cantor sets. One sees that a Cayley tree with $z_{(1)}=3$ and only one connected bond at each node cannot transmit a classical wave for more than two nodes. This observation was at the base of our conclusion in Ref. \cite{PRE14} that a nonlinear wave obeying NLSE~(\ref{1}) will be Anderson localized (similarly to a linear field), if $s < 1$. 

This conclusion, however, appears hasty somewhat in that it ignores the situations at the second and higher orders of multinomial expansion: 

Indeed, in the second order, one deals with a system of coupled nonlinear oscillators in Eq.~(\ref{Res2}), with the order-specific nonlinearity contained in the term $\beta_s \mathcal{{S}}_{k,(2)}$. This nonlinearity amends the nonlinearity at the first order [{\it i.e.}, the term $\beta_s \mathcal{{S}}_{k,(1)}$] and corresponds to a quintic polynomial in Eq.~(\ref{Delta}). As such, it is represented by a Cayley tree with the coordination number $z_{(2)}=5$ (see Fig.~2, right). In the corresponding mapping space one draws one integer bond for $\sigma_{m_3}$ and a Cantor set for each $\sigma^{q_{i_1,j_1}}_{i_1}$, $\sigma_{j_1}^{*q_{i_1,j_1}}$, $\sigma^{q_{i_2,j_2}}_{i_2}$, and $\sigma_{j_2}^{*q_{i_2,j_2}}$, suggesting, at first glance, that there is no asymptotic transport of the nonlinear field for precisely the same reasons as in the first order. 

\subsection{The homogeneity paradox (and its solution)}

It is at this point, however, where a subtlety occurs, and it refers to an observation that the original nonlinearity in Eq.~(\ref{2s}) corresponds to a {\it homogeneous} wave process in {all} orders. This, together with the fact that the coordination number of a Cayley tree is a topological invariant of the tree \cite{Schroeder}, would imply that the corresponding multinomial expansion of the nonlinear term $|\psi_n|^{2s}$ must refer to a Cayley tree with the {\it same} coordination number in {\it all} orders. 

There is an apparent paradox in this reasoning, and this is solved by demanding that some of the disrupted wave processes pertaining to the nonlinearity in Eq.~(\ref{Delta}) must couple together to form joint (degenerate) states, so the topological invariant $z=3$ is preserved in the second order of multinomial expansion. Then such states must be represented by a shared bond, and not by the proper bonds. We associate this coupling process with constructive interference among the disrupted waves. 

A closer inspection of dynamical Eq.~(\ref{Res2}) shows that at each node $k$ there must exist degenerate states with the multiplicity (degree of degeneracy) 3, corresponding to coupling among 3 (out of the 4 possible) disrupted waves with the amplitudes $\sigma^{q_{i_1,j_1}}_{i_1}$, $\sigma_{j_1}^{*q_{i_1,j_1}}$ and $\sigma^{q_{i_2,j_2}}_{i_2}$ and other combinations alike. 

It is understood that the constituent wave processes comprising degenerate states must be identical, so they will be characterized by the {\it same} power exponents $q_{i_1,j_1}$ or $q_{i_2,j_2}$. Because, on the other hand, $q_{i_1,j_1} + q_{i_2,j_2} = s$ in accordance with Eq.~(\ref{Sums}), one has $q_{i_1,j_1} = q_{i_2,j_2} = s/2$, from which the fractal dimension of degenerate states is found to be $d_f = 3s/2$. 

To this end, if $s \geq 2/3$, then $d_f \geq 1$, suggesting that the bonds representing degenerate states will be connected, if the exponent $s$ is large enough ({\it i.e.}, $s \geq 2/3$), and as such will be able to transmit a wave. 

One sees that there might occur asymptotic transport of the wave field already in the second order of multinomial expansion, if the exponent $s$ is greater than $2/3$. In the mapping space one draws one connected bond in place of the oval structure representing the degenerate state and ignores the remaining disrupted bond (see Fig.~3). Then the transport to long distances goes along alternating shared (degenerate) and proper (integer) bonds corresponding to the wave process $\sigma_{m_3} (t)$. 

\begin{figure}
\includegraphics[width=0.51\textwidth]{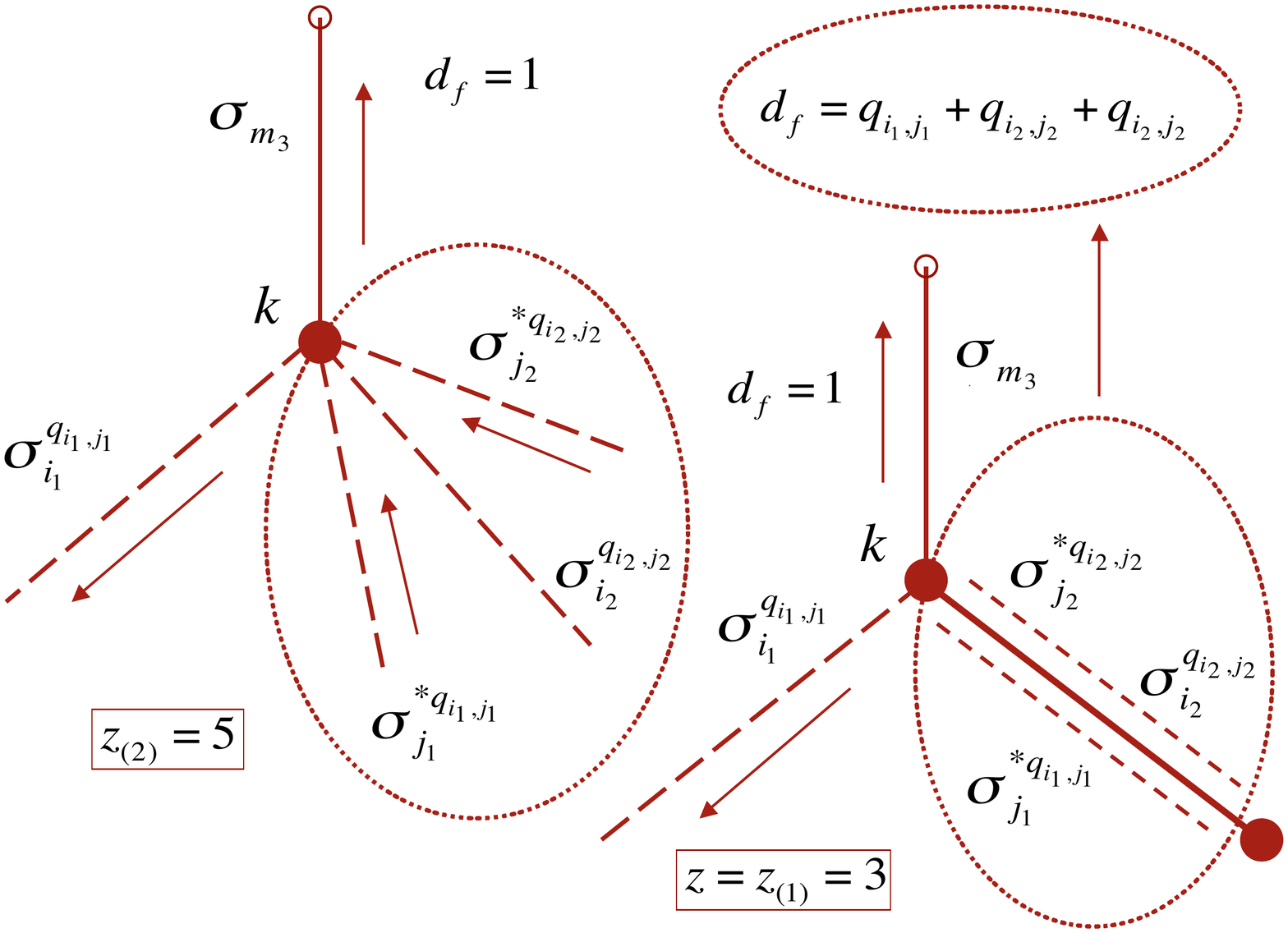}
\caption{\label{} Replacing disconnected bonds within a degenerate state (left) by a connected (wave transmitting) bond (right). The fractal dimension of the shared bond is $d_f = q_{i_1,j_1} + 2q_{i_2,j_2} = 3s/2$. The bond is conducting in the second order, if $d_f \geq 1$, demanding $s \geq 2/3$.
}
\end{figure}

\subsection{Self-intersecting Cayley trees}

From a topological perspective, the occurrence of degenerate states stems from the fact that in order to fold a Cayley tree {\it without} self-crossings in an Euclidean space, one needs a certain (exponentially large) number of embedding dimensions. This number depends on the coordination number of the tree \cite{Schroeder}. In that regard, when we assess the corrections from higher-order terms over the first-order term, we likewise are trying to embed a Cayley tree with the coordination number $z_{(2)}=5$ or higher into too narrow a space, which might be good to host a tree with the coordination number $z_{(1)}=3$ (see Fig.~2, left), though not more than this number. To this end, the embedding {\it without} self-crossings appears inappropriate; yet, one might allow for self-intersections of the tree. These self-intersections must, however, be such that the resulting Cayley trees at any higher order greater than one remain consistent with the topology of interactions in the primary order. Then the conservation of the topological invariant $z$ demands $z=3$, and is readily satisfied, if some of the disrupted states degenerate ({\it i.e.}, couple together to form a joint state). 

\begin{figure}
\includegraphics[width=0.51\textwidth]{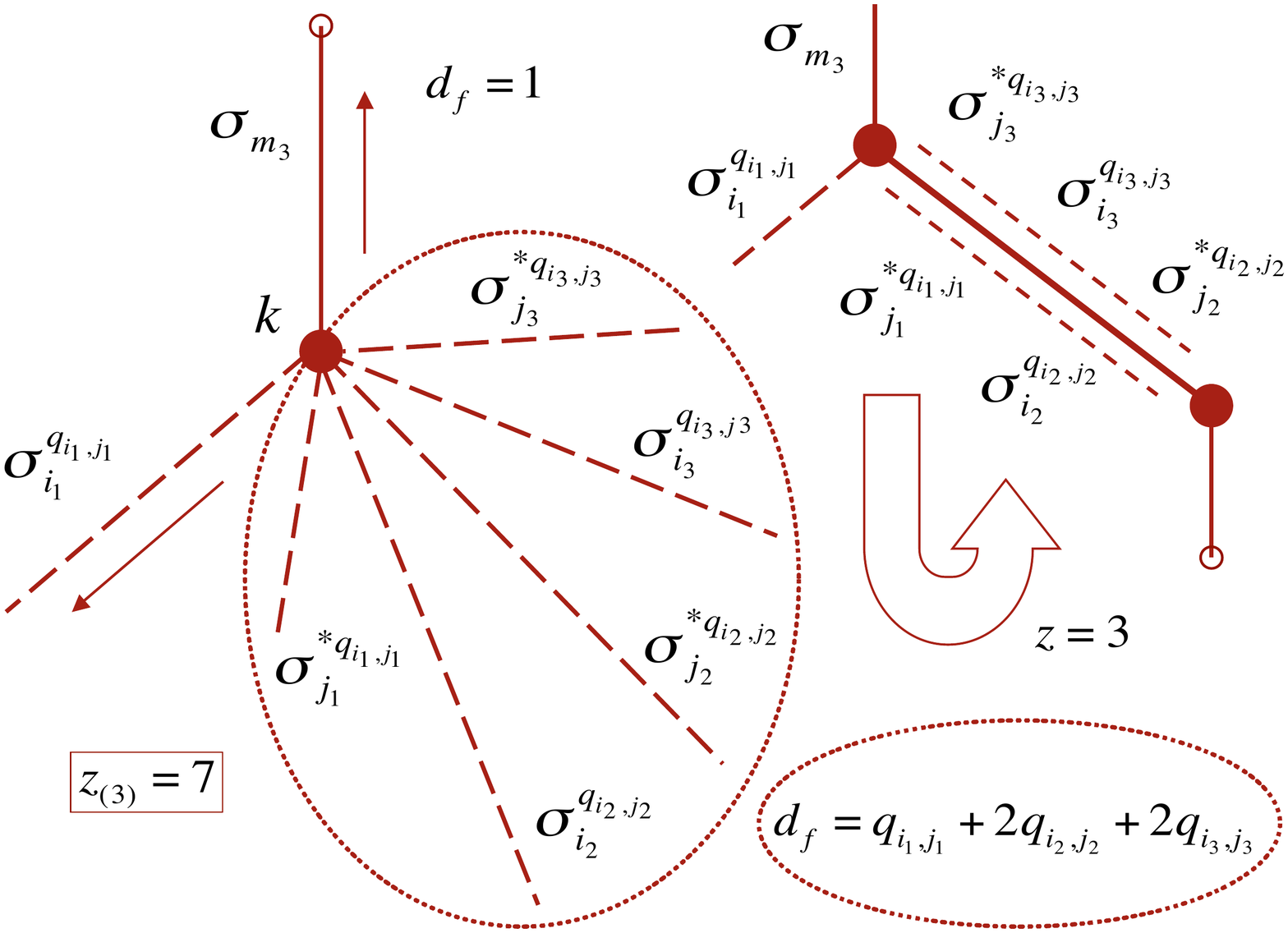}
\caption{\label{} Mapping dynamical equations for $\sigma_k (t)$ onto a Cayley tree in the third order. The wave processes giving rise to a degenerate state are inside the oval structure (dotted line). The reduction of these processes into a single (shared) bond is illustrated in the upper-right corner. The fractal dimension of this bond is given by the sum of respective fractional indices, i.e., $d_f = q_{i_1,j_1} + 2q_{i_2,j_2} + 2q_{i_3,j_3} = 5s/3$. The bond is conducting in the third order, if $d_f \geq 1$, demanding $s \geq 3/5$.
}
\end{figure}

\subsection{Mapping procedure in the third order}

Turning to the third order of multinomial expansion, one assumes that the condition in Eq.~(\ref{Sums}) is satisfied for three nonnegative (to ensure good behavior in the infrared limit), fractional-valued exponents, which we shall denote as $q_{i_1,j_1}$, $q_{i_2,j_2}$, and $q_{i_3,j_3}$. One has, accordingly, $q_{i_1,j_1} + q_{i_2,j_2} + q_{i_3,j_3} = s$. Then similarly to the above one obtains a system of dynamical equations for $\sigma_k (t)$, with septic polynomial terms (we do not write these terms here). 

In the mapping space, such terms will correspond to a Cayley tree with the coordination number $z_{(3)}=7$ (see Fig.~4, left). At each node $k$, this tree will contain one proper (integer) bond standing for the wave process $\sigma_{m_3}$, and as many as 6 disconnected bonds corresponding to the disrupted processes $\sigma^{q_{i_1,j_1}}_{i_1}$, $\sigma_{j_1}^{*q_{i_1,j_1}}$, $\sigma^{q_{i_2,j_2}}_{i_2}$, $\sigma_{j_2}^{*q_{i_2,j_2}}$, $\sigma^{q_{i_3,j_3}}_{i_3}$, and $\sigma_{j_3}^{*q_{i_3,j_3}}$. Each disrupted process will be represented by a Cantor set with the fractal dimension $d_f$ being equal to the respective power exponent $q_{i_1,j_1}$, $q_{i_2,j_2}$ or $q_{i_3,j_3}$. Then the homogeneity of the functional form in Eq.~(\ref{2s}) would imply that there must occur degenerate states with the multiplicity 5, so the topological invariant $z = 3$ is preserved in the third order. 

Demanding, once over again, that degenerate states are formed by identical wave processes, one gets $q_{i_1,j_1} = q_{i_2,j_2}=q_{i_3,j_2} = s/3$. It is understood that degenerate states in the third order arise from constructive interference among 5 (out of 6 possible) disrupted waves and are represented by a shared node in the mapping space (see Fig.~4, right), with the Hausdorff dimension $d_f = 5s/3$. To this end, if one requires $d_f \geq 1$, one infers that there might occur a nonvanishing transport to long distances in the third order, if $s \geq 3/5$. 

\subsection{Arbitrary order}

Generalizing, in an arbitrary order $\ell$ of multinomial expansion one constructs a Cayley tree with the coordination number $z_{(\ell)}=2\ell + 1$, such that at each node of the tree there will be one proper (integer) bond and as many as $2\ell$ disconnected bonds corresponding to Cantor sets with the Hausdorff fractal dimension $s/\ell$ each. Then the conservation of the topological invariant $z=3$ in the order $\ell$ would imply that any $2\ell - 1$ (out of the $2\ell$ present) disrupted waves must form, at each node $k$, a joint (degenerate) state with the degree of degeneracy $2\ell - 1$. The shared bond, representing this type of state in the mapping space, will have the Hausdorff dimension $d_f = (2\ell - 1)s/\ell$, and will be able to transmit a wave process, if $s \geq \ell / (2\ell - 1)$. Letting $\ell\rightarrow+\infty$, one concludes that NLSE~(\ref{1}) allows for unlimited transport to long distances ({\it i.e.}, the Anderson state destroyed by nonlinear interactions), if 
\begin{equation}
s \geq \lim_{\ell\rightarrow+\infty} \ell / (2\ell - 1) = 1/2.
\label{Iff} 
\end{equation} 
The result in Eq.~(\ref{Iff}) is remarkable and shows that the asymptotic transport of classical waves in NLSE~(\ref{1}) might occur for any $s \geq 1/2$, but one needs to incorporate {\it all} orders of multinomial expansion of $|\psi_n|^{2s}$ including arbitrarily high orders. 
The situation is much different from the analogue quantum case \cite{PRE19}, where the asymptotic transport to long distances is shown to occur for $s \geq 1/2$ already in the first order of multinomial expansion as a result of quantum tunneling of the disrupted waves.


\section{Scaling theory of field spreading}

If the field is spread across a large number of states $\Delta n \gg 1$, then the conservation of the total probability $\sum_n |\psi_n|^2 = 1$ demands that the density of the probability is small, {\it i.e.}, $|\psi_n|^2 \sim 1/\Delta n$. It is understood that the excitation of each new eigenstate is a spreading of the wave field in wave number space. The key point here is that the spreading process is mediated by resonances among the different oscillators participating in the interaction Hamiltonian~(\ref{6s+}). In fact, from Eq.~(\ref{6s+}), the  
resonance condition is  
\begin{equation}
\omega_k = s \omega_{m_1} - s \omega_{m_2} + \omega_{m_3}
\label{Res} 
\end{equation}
and involves a parametric dependence on the subquadratic power $s$ (due to the presence of disrupted wave processes). For $s = 0$ (linear model), the resonance condition in Eq.~(\ref{Res}) reduces to $\omega_k = \omega_{m_3}$ and is trivial, while for $s = 1$ (quadratic nonlinearity) it leads directly to Eq.~(10) of Ref. \cite{EPL}, yielding 
\begin{equation}
\omega_k = \omega_{m_1} - \omega_{m_2} + \omega_{m_3}.
\label{Res1} 
\end{equation}

\subsection{Chirikov overlap parameter}

When the resonances happen to overlap, the phase trajectories start to switch from one resonance to another on essentially a random basis \cite{Chirikov,ZaslavskyUFN}, giving rise to a stochastic spreading of the nonlinear field along the coordinate $n$. The onset of stochastic motions in the system corresponds to a situation according to which the nonlinear frequency shift resulting from the interactions present becomes greater than the distance between resonances in wave number space. Focusing on NLSE~(\ref{1}), one sees that the nonlinear frequency shift behaves with the probability density as $\Delta\omega_{\rm NL} = \beta_s (|\psi_n|^{2})^s$ and for $\Delta n \gg 1$ will scale with the number of states in accordance with $\Delta\omega_{\rm NL} \simeq \beta_s / (\Delta n)^s$. On the other hand, with an increasing $\Delta n$, the distance between resonances, $\delta\omega$, goes to zero as $\propto 1/\Delta n$ consistently with the conditions in Eqs.~(\ref{Res}) and~(\ref{Res1}). The Chirikov overlap parameter $\mathcal{K} \simeq \Delta \omega_{\rm NL} / \delta\omega$ \cite{Chirikov} shows by how much the nonlinear frequency shift exceeds the distance $\delta \omega$ and in this sense measures the degree of chaoticity in the system \cite{Zaslavsky,Sagdeev} (in this paradigm, large $\mathcal{K}$ values correspond to strong chaos, {\it i.e.}, $\mathcal{K}\gg 1$, while the values like $\mathcal{K}\sim 1$ are borderline). 
Using $\Delta\omega_{\rm NL} \simeq \beta_s / (\Delta n)^s$ and $\delta\omega\simeq 1/\Delta n$, one gets, in the strong chaos regime, 
\begin{equation}
\mathcal{K} \simeq \beta_s (\Delta n)^{1-s} \gg 1.
\label{K-value} 
\end{equation}
It is noted that the $\mathcal{K}$ value depends, in general, on the number of states (except for the quadratic power case, for which $s=1$). For $s < 1$, the onset of field spreading may be sensitive to both $\beta_s$ and initial spread. If, however, the initial spread is such that the condition $\mathcal{K} \gg 1$ is satisfied at time $t=0$, then for $s < 1$ it will be satisfied for any $t > 0$, giving rise to ever pursuing stochastic motions in the system.  Moreover, the chaos will be self-reinforcing in that the $\mathcal{K}$ value increases with an increasing $\Delta n$. On the other hand, it is clear that the Chirikov criterion in Eq.~(\ref{K-value}) is necessary, though not yet sufficient, a condition for the stochastic spreading to come into play (because the chaotic motions may be Anderson localized by the inhomogeneities present \cite{Sh93,PS,Flach}). If, however, $s \geq 1/2$, then the localization is destroyed by nonlinear interaction in accordance with Eq.~(\ref{Iff}) above, enabling unlimited spreading of the wave field to large distances along the domains of chaotic dynamics. 

\subsection{Dynamical model}

In the parameter range of strong chaos, the rate of field spreading $R = d\Delta n / dt$ is obtained using NLSE~(\ref{1}) to give, with the aid of $|\psi_n|^2 \sim 1/\Delta n$, 
\begin{equation}
R\sim |\dot{\psi_n}|^2 \sim \beta_s^2\, \big||\psi_n|^{2s} \psi_n\big|^2 \sim A/(\Delta n)^{2s+1},
\label{Seq} 
\end{equation}
leading to
\begin{equation}
d\Delta n / dt = A/(\Delta n)^{2s+1},
\label{RR} 
\end{equation}
where $\dot{\psi_n}$ denotes the time derivative of $\psi_n$, and $A\propto\beta_s^2$ is a numerical coefficient. In writing Eq.~(\ref{Seq}) we have assumed that the phases of comprising wave processes are essentially random (that is, eventual correlation length is much smaller than $\Delta n$). Integrating over time in Eq.~(\ref{RR}), one gets $(\Delta n)^{2s + 2} = (2s + 2) A t$, from which a sub\-diffusive spreading law    
\begin{equation}
(\Delta n)^2 = [(2s + 2) A]^{1/(s+1)}\,t^{1/(s+1)}
\label{Sub} 
\end{equation}
can be deduced for $t\rightarrow+\infty$. For $s\rightarrow 1$, one finds $(\Delta n)^2 \propto t^{1/2}$ consistently with the numerical result of Ref. \cite{Skokos18} in the strong chaos regime. Remark that the quadratic nonlinearity, characterized by $s=1$ exactly, is special in that it allows for both the strong and weak chaos regimes \cite{EPL,PRE14}. In the latter case, the scaling law $(\Delta n)^2 \propto t^{1/3}$ is found (see below). 

Differentiating the both sides of Eq.~(\ref{RR}) with respect to time and eliminating the remaining $d\Delta n / dt$ with the aid of same Eq.~(\ref{RR}), one gets 
\begin{equation}
d^2 \Delta n /dt^2 = -(2s + 1)A^2 / (\Delta n)^{4s + 3}.
\label{Grad+} 
\end{equation}
Using a gradient form on the right-hand side of Eq.~(\ref{Grad+}), one obtains
\begin{equation}
\frac{d^2}{dt^2}\Delta n = - \frac{d}{d \Delta n} \left[- \frac{A^2 / 2}{(\Delta n)^{4s + 2}}\right].
\label{Grad} 
\end{equation}
Equation~(\ref{Grad}) is none other than a Newtonian equation of motion along the ``coordinate" $\Delta n$ in the potential field 
\begin{equation}
W (\Delta n) = - \frac{A^2 / 2}{(\Delta n)^{4s + 2}}.
\label{Poten} 
\end{equation}
For $s\rightarrow 1$, the potential function in Eq.~(\ref{Poten}) becomes $W(\Delta n) = - (A^2 / 2)/ (\Delta n)^6$, and is easily seen to be the attractive part of the celebrated Lennard-Jones potential \cite{Lennard}, known from molecular physics. Given the attracting character of $W (\Delta n)$, one might arguably propose that the newly excited modes would tend to form clusters (``molecules") in wave number space; where, they will be effectively trapped \cite{Iomin} due to their nonlinear coupling.    

Multiplying the both sides of Eq.~(\ref{Grad}) by the ``velocity", $d \Delta n /dt$, and integrating the ensuing differential equation with respect to time, after a simple algebra one obtains
\begin{equation}
\frac{1}{2}\left[\frac{d}{dt} \Delta n\right]^2 - \frac{A^2 / 2}{(\Delta n)^{4s + 2}} = \Delta E,
\label{Ener} 
\end{equation}
where the first term on the left-hand side has the sense of the kinetic energy of a ``particle" of unit mass moving along the coordinate $\Delta n$, and the second term is its potential energy. It is shown using Eq.~(\ref{RR}) 
that the kinetic energy in Eq.~(\ref{Ener}) compensates for the potential energy {\it exactly}, that is, the full energy in Eq.~(\ref{Ener}) is zero, {\it i.e.}, $\Delta E = 0$. More so, both the negative potential energy $W (\Delta n) \sim - A^2 / 2(\Delta n)^{4s + 2}$ and the positive kinetic energy $\frac{1}{2}(d \Delta n /dt)^2 \sim A^2 / 2(\Delta n)^{2(2s + 1)}$ vanish while spreading. Both will decay as the $(4s + 2)$-th power of the number of states, and the ratio between them will {\it not} depend on the width of field distribution. 

The full energy being equal to zero implies that the ``particle" in Eq.~(\ref{Ener}) is sitting on the separatrix $\Delta E = 0$. Based on the analysis of Ref. \cite{PRE00}, one might demonstrate that the separatrix $\Delta E = 0$ contains a connected escape path to infinity, hence allows for an unlimited spreading of the wave field regardless of how large is $\beta_s$ (for $s < 1$; the case $s=1$ appears special and should be taken with more care: see Sec.~VIII). 

More so, as the particle propagates outward, its motion becomes intrinsically unstable (sensitive to fluctuations). This is due to the peculiar character of separatrix transport \cite{Zaslavsky,Sagdeev}, implying that tiny perturbations due to, for instance, random noise or imprecision in the initial conditions may drastically change the type of phase space trajectory. The observation is especially relevant for separatrix dynamics in large systems \cite{ChV}. At this point, the fact that a given mode does or does not belong to a given cluster of modes becomes essentially a matter of the probability. 

\subsection{Waiting-time distribution}

To assess the probabilistic aspects of field spreading, let us assume that the fluctuation background in Eq.~(\ref{Ener}) is characterized by a thermodynamic ``temperature," $T$. That is, the value of $T$ weighs any occasional perturbations to dynamics that might be influential near the separatrix. Then the probability for a given mode to quit the cluster after it has traveled $\Delta n$ sites on it could be written as the Boltzmann factor 
\begin{equation}
p (\Delta n) = \exp [W (\Delta n) / T].
\label{BF} 
\end{equation}
Substituting $W (\Delta n)$ from the potential function in Eq.~(\ref{Poten}), one finds
\begin{equation}
p (\Delta n) = \exp [- A^2 / 2 T (\Delta n)^{4s + 2}].
\label{Escape} 
\end{equation}
Taylor expanding the exponential function for $\Delta n \gg 1$, one gets 
\begin{equation}
p (\Delta n) \simeq 1 - A^2 / 2 T (\Delta n)^{4s + 2}.
\label{Expand} 
\end{equation}
The probability to remain (``survive") on the cluster after $\Delta n$ space steps is $p^{\prime} (\Delta n) = 1-p(\Delta n)$, yielding 
\begin{equation}
p^{\prime} (\Delta n) \simeq A^2 / 2 T (\Delta n)^{4s + 2}.
\label{EscapePr} 
\end{equation}
Eliminating $\Delta n$ with the aid of Eq.~(\ref{Sub}), one obtains the probability to survive on the cluster after $\Delta t$ time steps  
\begin{equation}
p^{\prime} (\Delta t) \propto (\Delta t)^{-(2s+1) / (s+1)},
\label{Survive} 
\end{equation}
leading to a waiting-time distribution \cite{Klafter,Sokolov} 
\begin{equation}
\chi_\alpha (\Delta t) \propto (\Delta t)^{-(1+\alpha)}
\label{WT} 
\end{equation}
with $\alpha = s/(s+1) < 1$. We associate the distribution in Eq.~(\ref{WT}) with the binding effect of finite clusters \cite{Iomin}. Note that the integral 
\begin{equation}
\int_{\sim 1}^{\tau} \Delta t \chi_\alpha (\Delta t) d\Delta t \sim \int_{\sim 1}^{\tau} (\Delta t)^{-\alpha} d\Delta t \sim \tau^{1-\alpha} \rightarrow+\infty
\label{Div} 
\end{equation}
diverges for $\tau\rightarrow+\infty$, implying that the mean waiting time is infinite. 

\section{Kinetic theory}

Let us now obtain a kinetic equation for asymptotic spreading. For this, we adopt the theoretical scheme of continuous time random walks (CTRW) \cite{CTRW1,CTRW2,Bouchaud}, according to which the transport occurs as a result of random-walk jumps along the coordinate $n$ with a distribution of waiting times between consecutive steps of the motion. Combining~(\ref{WT}) with a simplifying assumption (to be revisited below) that there is a characteristic jump length of the random process, one arrives at a non-Markovian generalization of the diffusion equation \cite{Klafter,Sokolov} 
\begin{equation}
\frac{\partial}{\partial t} f (n, t) =\,_{0}\mathrm{D}_t^{1-\alpha}\Big(K_{\alpha} \frac{\partial^{2}}{\partial n^{2}} f (n, t)\Big).
\label{FDEL-L} 
\end{equation}
Here, $f = f (n, t)$ is the probability density to find the random walker at time $t$ at the distance $n$ away from the origin, $K_{\alpha}$ is the transport coefficient and carries the dimension cm$^{2}\,\cdot\,$s$^{-\alpha}$, and 
\begin{equation}
{_0}\mathrm{D}_t^{1-\alpha} f (n, t) = \frac{1}{\Gamma (\alpha)}\frac{\partial}{\partial t}\int _{0}^{t} \frac{f (n, t^{\prime})}{(t - t^{\prime})^{1-\alpha}}dt^\prime \label{R-L} 
\end{equation}
is the so-called Riemann-Liouville fractional derivative \cite{Podlubny,Samko} of order $1-\alpha$, which incorporates the trapping effect of the clusters consistently with the waiting-time distribution in Eq.~(\ref{WT}). Note that we directly associate the non-Markovian character of Eq.~(\ref{FDEL-L}) with the divergence of mean waiting time in Eq.~(\ref{Div}). Based on kinetic Eq.~(\ref{FDEL-L}), one finds the asymptotic ($t\rightarrow+\infty$) mean-square displacement of the random walker to be    
\begin{equation}
\langle (\Delta n)^2 (t) \rangle \propto t^{\,\alpha},
\label{MSD} 
\end{equation} 
where $\alpha = s/(s+1)$. Because $\alpha < 1$, the spreading process is subdiffusive. Comparing to Eq.~(\ref{Sub}), one sees that the two processes are consistent, if (and only if) $s=1$, while for $s < 1$ there is a discrepancy to be repaired in some way. This observation suggests that the non-Markovian kinetic equation~(\ref{FDEL-L}) is good for NLSE~(\ref{QNLSE}), with quadratic power nonlinearity, though {\it not} for the generalized model in Eq.~(\ref{1}), with $s < 1$. 

To remedy, let us assume that the trapping process in Eq.~(\ref{WT}) competes with the possibility for the random walker to perform long-distance jumps along the coordinate $n$, with a power-law distribution of jump lengths 
\begin{equation}
\chi_\mu (|\Delta n|) \propto |\Delta n|^{-(1+\mu)},
\label{LW} 
\end{equation}    
where $\mu$ is a power exponent. We associate the distribution in Eq.~(\ref{LW}) with occasional jumps between the different clusters in wave number space. 

From a kinetic perspective, the effect such jumps would have on the transport Eq.~(\ref{FDEL-L}) is that the Laplacian operator ${\partial^{2}}/\partial {n^{2}}$ must be replaced by the integro-differential operator 
\begin{equation}
\frac{\partial^\mu}{\partial |n|^\mu} f (n, t) = \frac{1}{\Gamma_\mu}\frac{\partial^2}{\partial n^2} \int_{-\infty}^{+\infty}\frac{f (n^\prime, t)}{|n-n^\prime|^{\mu - 1}} dn^\prime,
\label{Def+} 
\end{equation} 
leading to a generalized kinetic equation of the form \cite{Klafter,Rest}
\begin{equation}
\frac{\partial}{\partial t} f (n, t) =\,_{0}\mathrm{D}_t^{1-\alpha}\Big(K^{\mu}_{\alpha} \frac{\partial^{\mu}}{\partial |n|^{\mu}} f (n, t)\Big).
\label{FDEL} 
\end{equation}
In the above, $\Gamma_\mu = -2\cos(\pi\mu/2)\Gamma(2-\mu)$ is a normalization parameter, which is introduced to ensure smooth crossover to the Laplacian operator in the limit $\mu\rightarrow 2$; $K^{\mu}_{\alpha} $ is the generalized transport coefficient, which, in the present case, carries the dimension cm$^{\mu}\,\cdot\,$s$^{-\alpha}$; and we have tacitly assumed that the exponent $\mu$ lies within $1 < \mu < 2$. The latter assumption guarantees that the power-law distribution in Eq.~(\ref{LW}) belongs to a class of L\'evy stable distributions \cite{Bouchaud,Gnedenko}. The interval $0 < \mu < 1$, although similar, is not considered here (not relevant for our purposes). 

Mathematical (as well as statistical-mechanical) foundations of the transport Eq.~(\ref{FDEL}) are spelled out in Refs. \cite{Klafter,Sokolov,Rest}. In a nutshell, this equation describes a random-walk process with competition between trappings ($\alpha < 1$) and jumps ($\mu < 2$), and has been found in a number of dispersive systems with disorder ({\it e.g.}, Refs. \cite{Klafter,Rest} for reviews; references therein). We should stress that the inclusion of long-haul jumps via the power-law distribution in Eq.~(\ref{LW}) leads to a very special form of the spatial derivative, Eq.~(\ref{Def+}), which includes a nonlocal integration over $n^\prime$ [via the convolution with the probability density $f(n^\prime, t)$]. That nonlocality would naturally arise in our model might be expected from NLSE~(\ref{DSNL}) with distributed nonlinearity, where the long-range dependence is introduced via a convolution of the order parameter $|\psi_n^\prime|^2$ with a slowly decaying response function, $\chi (n-n^\prime) \propto 1/|n-n^\prime|^s$. In a basic theory of random processes, the nonlocal operator in Eq.~(\ref{Def+}) is referred to as the Riesz fractional derivative \cite{Klafter,Rest}. As is well known \cite{Klafter,Ch2007}, this type of derivative generates L\'evy flights. For $\alpha \rightarrow 1$ and $\mu\rightarrow 2$, the transport process in Eq.~(\ref{FDEL}) transforms into a normal (Gaussian) diffusion along $n$ (because ${\partial^{\mu}}/\partial {|n|^{\mu}} \rightarrow {\partial^{2}}/\partial {n^{2}}$ in that case). For $\mu\rightarrow 1$, the exact analytical representation of ${\partial^{\mu}}/\partial {|n|^{\mu}}$ is obtained via a reduction of Eq.~(\ref{Def+}) to the Hilbert transform operator \cite{Mainardi}, yielding, instead of~(\ref{Def+}), 
\begin{equation}
\frac{\partial^\mu}{\partial |n|^\mu} f (n, t)  = - \frac{1}{\pi} \frac{\partial}{\partial n} \int_{-\infty}^{+\infty}\frac{f (n^\prime, t)}{n-n^\prime} dn^\prime, \ \ \ \mu=1.
\label{Hilbert} 
\end{equation} 
Using~(\ref{FDEL}), one obtains the fractional moments \cite{Klafter} of the $f = f (n, t)$ distribution, from which the scaling of the pseudo-mean square displacement may be deduced for $t\rightarrow+\infty$, leading to 
\begin{equation}
\langle (\Delta n)^2 (t) \rangle = \lim_{\delta\rightarrow 2}\,(\Delta n)^\delta \propto t^{2\alpha/\mu}.
\label{Pseudo} 
\end{equation} 
An exact calculation of the fractional moments of $f (n, t)$ uses the formalism of the Fox $H$-functions \cite{Podlubny,Samko} and is articulated in Refs. \cite{Klafter,Rest,Ch2007}. Comparing to Eq.~(\ref{Sub}), one infers $2\alpha/\mu = 1/(1+s)$, from which $\mu = 2s < 2$, where the general relation $\alpha = s/(s+1)$ stemming from Eq.~(\ref{WT}) was applied. 

One sees that the generalized kinetic equation in Eq.~(\ref{FDEL}) is the right tool to describe the complex transport processes pertaining to NLSE~(\ref{1}). 

Perhaps of greater importance is the fact that the exponent $s$ being smaller than 1 implies $\mu < 2$, showing that the transport is {\it always} nonlocal, with L\'evy flights, if the power $s$ is sublinear ({\it i.e.}, nonlinearity at the NLSE is subquadratic). 

From a topological perspective, the fact that we have encountered L\'evy flights for $s < 1$ is clear from the iteration procedure of Sec.~V, according to which the inclusion of subquadratic power results in the occurrence of multiple degenerate states in wave number space. In that interpretation, nonlocality arises as a consequence of improper embedding of the higher-order Cayley trees into the primary order mapping space, giving rise to L\'evy flights along the shared (degenerate) bonds. In the limit of quadratic power nonlinearity, {\it i.e.}, $s \rightarrow 1$, the iteration procedure of Sec.~V is exact in the first order. That means that nonlocality is lost for $s \rightarrow 1$, implying that the asymptotic spreading of the nonlinear field goes on next-neighbor transitions between adjacent states, with a subdiffusive dispersion    
\begin{equation}
\langle (\Delta n)^2 (t) \rangle \propto t^{1/2}.
\label{MeanS} 
\end{equation} 
This type of subdiffusive process with long-time trapping phenomena has been observed numerically in Ref. \cite{Skokos18} in the regime of strong chaos. Exactly the same behavior was found in Ref. \cite{PRE17} for quantum nonlinear Schr\"odinger lattices with randomness; where, it was associated with the binding effect of the Lennard-Jones potential, $W (\Delta n) = - (A^2 / 2)/(\Delta n)^{6}$. In that regard, we note that the dynamical picture of quantum chaos may be much different from the classical one in that it is based on Fermi's golden rule for transitions between states \cite{PRE19}, with a greater emphasis on nonlocal features through the transport.

\section{Why quadratic nonlinearity is special}

If $s=1$, then instead of NLSE~(\ref{1}) one relies on a simplified model in Eq.~(\ref{QNLSE}), with the quadratic nonlinearity $\propto \beta \,|\psi_n|^2$. In that case, the Chirikov overlap parameter $\mathcal{K} = \Delta\omega_{\rm NL}/\delta\omega \simeq \beta$ does not depend on the width of field distribution: If for some $\beta$ the field is chaotic satisfying $\mathcal{K} \gg 1$ at time $t=0$, then the actual $\mathcal{K}$ value will be preserved while spreading. Therefore, no self-reinforcing of chaos is expected for $s=1$ (which according to the above excludes L\'evy flights), suggesting that the asymptotic transport goes on random transitions between adjacent ({\it i.e.}, next-neighbor) states. With these implications in mind, one reinstalls the Laplacian operator $\partial^{2}/\partial n^{2}$ in kinetic Eq.~(\ref{FDEL}), leading to local Eq.~(\ref{FDEL-L}) despite that the chaos could be strong. If, however, the $\beta$ value is so small that the overlap parameter $\mathcal{K} \ll 1$ for $t = 0$, then the transition to chaos does not occur, and the field remains localized for all $t > 0$. Separating the two regimes (chaotic {\it vs.} localized) is the borderline regime, characterized by $\mathcal{K} \sim 1$, which might be expected for $s=1$ (for a suitably chosen $\beta$), but not really for $s < 1$ (because the dynamical chaos being self-reinforcing for $s < 1$ naturally destroys the borderline behavior in the long run). 

{\it A priori} one might predict that the borderline regime (which we associate with weak chaos considered in Refs. \cite{Report,Mod,PRE09}) is characterized by the non-Markovian transport equation~(\ref{FDEL-L}), though with a different $\alpha$ value [because the result $\alpha = s/(s+1) = 1/2$ has used that the chaos is strong, {\it i.e.}, the condition $\mathcal{K} \gg 1$ holds]. In what follows, we obtain a proper value of $\alpha$ for $\mathcal{K} \sim 1$.

\subsection{Threshold percolation on a Cayley tree}

The main idea here \cite{EPL} is that the transport with $\mathcal{K} \sim 1$ occurs in the form of next-neighbor jumps along a system of dephased oscillators, thought of as ``conducting" (wave transmitting) sites for the spreading process. Of interest here is a situation when such a system could stretch to infinitely long scales, so it supports unlimited spreading of the nonlinear field in wave number space. The existence of such a system is not at all obvious. If, however, the concentration of dephased oscillators exceeds a certain threshold value, then a connected cluster of dephased oscillators occurs with the probability 1, thus offering a substrate for unlimited spreading. As is shown in a basic theory of random processes, this substrate corresponds to the infinite percolation cluster at the edge of percolation \cite{Naka,Stauffer,Isi}. 

More explicitly, we divide all oscillators in Eq.~(\ref{4}) into two categories, {\it i.e.}, those ``dephased" for which the local $\mathcal{K}$ value is large enough enabling random transitions between adjacent states, and those in a ``regular" state for which the local $\mathcal{K}$ value is so small that no transition to a neighboring state may occur. Next, let us assume, following the approach of Refs. \cite{EPL,PRE14}, that each nonlinear oscillator resides in a dephased state with the probability $p$, and in a regular state with the probability $1-p$. In this fashion, one mimics the formulation of the random percolation problem on a lattice \cite{Stauffer,Isi}, the result being that there exists a critical value of $p$ for which an infinite connected cluster of dephased oscillators occurs for the first time. This value, which depends \cite{Naka,Feder} on the topology of the lattice on which the percolation transition is analyzed, is none other than the percolation threshold in the system of coupled nonlinear oscillators in Eq.~(\ref{4}). We associate this critical value with the onset of unlimited transport in the weak chaos regime. 

Next, we need to discuss what a suitable lattice for the percolation transport would be in the framework of NLSE~(\ref{QNLSE}). This is clear from the mapping procedure articulated in the beginning of Sec.~V and the fact that the multinomial expansion of quadratic nonlinearity is exact in the first order. At this point, one is led to conclude that the percolation occurs on a very specific lattice, which is provided by a Cayley graph with the coordination number $z=3$ (see Fig.~5). We note in passing that this type of lattice offers a convenient model of mean-field percolation \cite{Naka,Schroeder}. 

Finally, let us consider that the transitions between dephased oscillators are entirely random and follow next-neighbor communications between the adjacent states. The random character of the transitions is motivated by the fact that the nonlinear field is chaotic along the percolation clusters (and is regular otherwise). Note, however, that the phase space available for random dynamics may be actually quite narrow at the edge of percolation, being reduced to a set of network-like structures \cite{PRE00} that are geometrically complex and strongly shaped. Note, also, that the percolation clusters, whatever finite or infinite, occupy only a fraction of the ambient space and {\it not} the entire space. The implication is that the distribution of domains of chaotic motions is highly inhomogeneous for $\mathcal{K} \sim 1$, at contrast to the strong chaos regime, with $\mathcal{K} \gg 1$, for which the formation of a wide ``stochastic sea" of chaotic dynamics is the case \cite{Zaslavsky,Sagdeev,Report}. 

\subsection{Random-walk model}

Based on the discussion above, we might arguably propose that a suitable kinetic model to characterize the onset spreading of the nonlinear field could be obtained as a random-walk model on a mean-field percolation cluster (by which one means an infinite percolation cluster on the Cayley graph \cite{Naka,Schroeder}) (see Fig.~5). This transport problem has been well and widely studied in the literature ({\it e.g.}, Refs. \cite{Naka,Havlin} for reviews; references therein) and is shown to result in a subdiffusive dispersion of the random walker in accordance with \cite{Gefen}    
\begin{equation}
\langle (\Delta n)^2 (t) \rangle \propto t^{2/(2+\theta)},
\label{Percol} 
\end{equation} 
where $\theta$ is the so-called connectivity index, which is a topological invariant of the cluster \cite{UFN,PRE97}. The latter index characterizes the deviation from normal diffusion in fractal geometry due to the presence of cycles, voids and dead-ends throughout the fractal distribution. For mean-field percolation on trees, $\theta = 4$ (Ref. \cite{Naka}; references therein; a derivation using the graph theory is presented in Ref. \cite{DNC}), leading, in view of Eq.~(\ref{Percol}), to the scaling behavior
\begin{equation}
\langle (\Delta n)^2 (t) \rangle \propto t^{1/3}.
\label{Percol2} 
\end{equation} 
The scaling law in Eq.~(\ref{Percol2}) will be consistent with kinetic Eq.~(\ref{FDEL-L}), if $\alpha = 1/3$. Note that the result $\alpha = s/(s+1)$ pertaining to strong chaos does not apply here \cite{Comm2}. One sees that the borderline transport, with $\mathcal{K} \sim 1$, leads to a slower spreading process as compared to the strong chaos regime, characterized by $\alpha = 1/2$. These theory predictions find support in computer simulations of Ref. \cite{Skokos18}, where both the strong and weak chaos regimes have been recognized numerically, with the respective transport exponents $\alpha = 1/2$ and $\alpha = 1/3$.   

\begin{figure}
\includegraphics[width=0.51\textwidth]{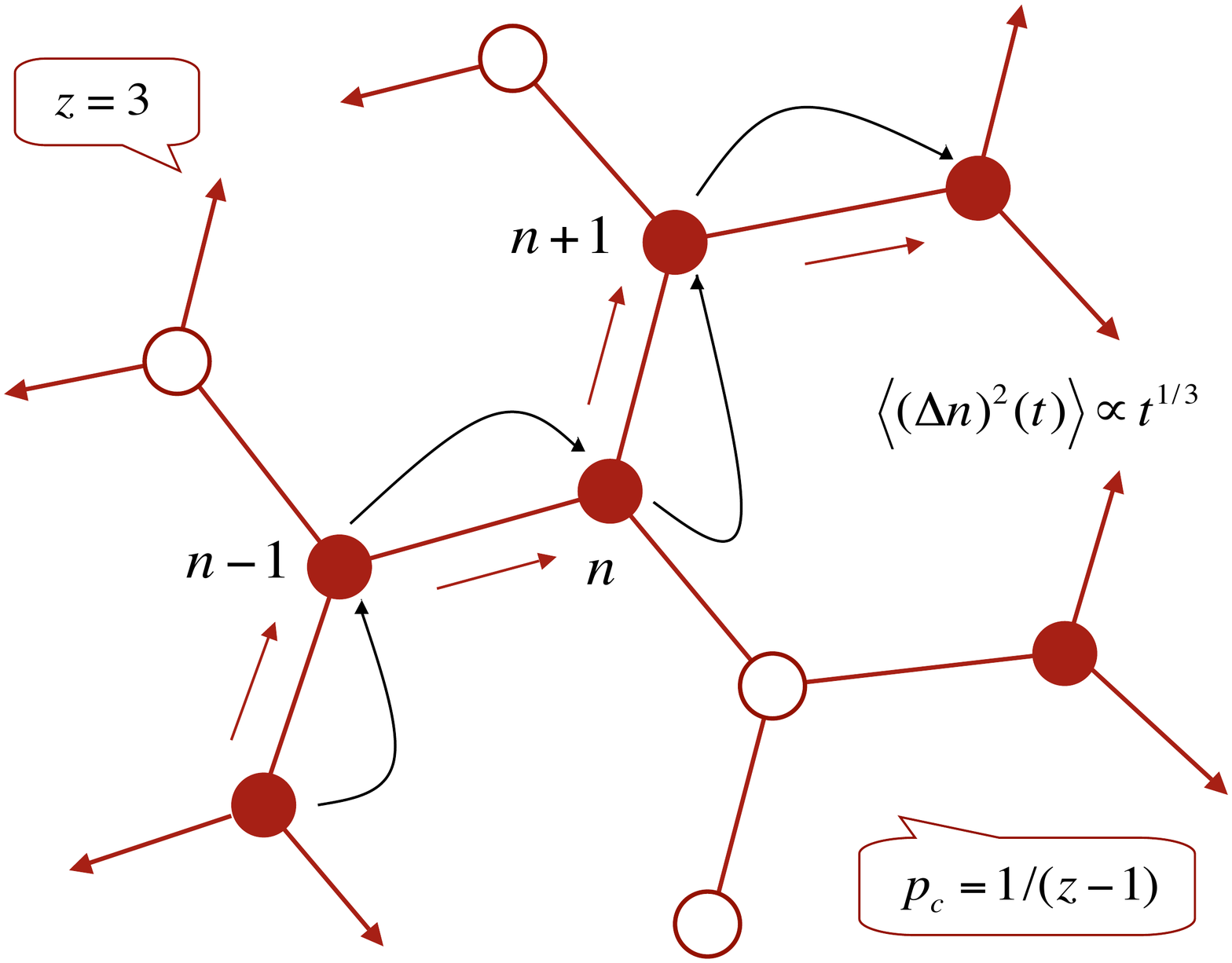}
\caption{\label{} Next-neighbor random walk on a percolating system of dephased oscillators (filled circles). The trajectory of the random walk is represented by a piecewise curve (black color). The oscillators in regular state (not available for the random walk) are shown as empty circles. It is assumed that dephased oscillators are organized in connected infinite ({\it i.e.}, ``percolating") clusters lying on a Cayley graph with the coordination number $z=3$. The onset of unlimited transport corresponds to a critical concentration of dephased oscillators, which depends on the coordination number $z$ and is given by $p_c = 1/(z-1)$. The dispersion of the random walk corresponds to the scaling $\langle (\Delta n)^2 (t) \rangle \propto t^{1/3}$ for $t\rightarrow+\infty$.   
}
\end{figure}

\subsection{Critical $\beta$ value}

Here we obtain the critical value of $\beta$, for which an unlimited transport along the lattice occurs for the first time, provided just that the nonlinearity is quadratic, {\it i.e.}, $s=1$. As is already clear, this critical value corresponds to the onset of percolation on a Cayley tree with the coordination number $z=3$. In a basic theory of threshold percolation on trees it is shown \cite{Naka,Schroeder} that the onset point arises for a critical probability of site occupancy, which depends on the coordination number $z$ (and only on this number). This critical probability is given by \cite{Schroeder}
\begin{equation}
p_c = 1/(z-1).
\label{Cri} 
\end{equation} 
Setting $z=3$, one readily gets $p_c = 1/2$. In the framework of NLSE~(\ref{QNLSE}), the threshold probability in Eq.~(\ref{Cri}) corresponds to a critical concentration of dephased oscillators, along which a weakly chaotic nonlinear field can propagate to long distances. Writing the concentration as the Boltzmann factor
\begin{equation}
p = \exp (-\delta\omega/\Delta\omega_{\rm NL}) = \exp (-1/\beta),
\label{Exp} 
\end{equation} 
we have, at the percolation point, $p_c = \exp (-1/\beta_c)$, where $\beta_c$ denotes the critical $\beta$ value, and we have considered that the nonlinear frequency shift, $\Delta\omega_{\rm NL} = \beta |\psi_n|^2$, plays the role of effective ``temperature" of nonlinear interaction \cite{EPL,PRE14}. Note that the behavior in Eq.~(\ref{Exp}) is non-perturbative (as, in fact, it must be in vicinity of a critical point \cite{Sornette2004}). Applying~(\ref{Cri}), one finds, with $z=3$, 
\begin{equation}
\beta_c = 1/\ln\, (z - 1) = 1/\ln 2 \approx 1.442695\dots
\label{Beta} 
\end{equation} 
The latter value, which is an exact result of the percolation model, defines the critical strength of nonlinear interaction, such that above this strength the nonlinear field in Eq.~(\ref{QNLSE}) can propagate along the lattice to unlimited distances, and is Anderson localized similarly to a linear field otherwise. Note that the onset of unlimited spreading is a thresholded ({\it i.e.}, critical) phenomenon, as it requires the strength of nonlinear interaction to exceed a certain finite value. We should stress that the existence of such a value is a property of quadratic nonlinearity, with $s=1$. If $s < 1$, then the respective probability of site occupancy
\begin{equation}
p = \exp (-\delta\omega/\Delta\omega_{\rm NL}) = \exp [-1/\beta_s (\Delta n)^{1-s}]
\label{Exp-s} 
\end{equation} 
depends on $\Delta n$, and for $\Delta n \rightarrow \infty$ will be (promptly) converging to $1$, implying that the field is chaotic. Thus, no percolation-like transport could be expected for $s < 1$, suggesting that the quadratic nonlinearity is special in that it is the only type of nonlinearity, which supports both the strong and weak transport regimes consistently with the conclusion of Ref. \cite{PRE14}. 
   
\section{Summary} 

Our analysis indicates that the nonlinear Schr\"odinger models with random potential and subquadratic power nonlinearity might constitute an efficient and powerful tool when describing dynamical systems with competition between nonlinearity, disorder and long-range dependence. In this respect, a very convenient issue is that the subquadratic nonlinearity in NLSE~(\ref{1}) arises as a reduction of the distributed nonlinearity with a power-law response function, $\chi (n-n^\prime) \propto 1/|n-n^\prime|^s$. This observation suggests that NLSE~(\ref{1}) may be directly applied to complex systems at or near their critical states \cite{Sornette2004,Pruessner,Asch2013}. Mathematically, our approach is different from the alternative approach articulated in Refs. \cite{Weitzner,PLA2005,Korabel2007}, where the long-range dependence is introduced into the linear (dispersion) term, leading to a L\'evy-fractional NLSE \cite{Iomin21,UFN}. 

In this paper's work, the main emphasis has been on the analytical methods pertaining to subquadratic power nonlinearity and a way to tackle this type of nonlinearity in the context of NLSE~(\ref{1}). In that regard, we have devised an iteration procedure based on Diophantine equations and the multinomial theorem, according to which one can represent the subquadratic power as a sequence of the Cayley graphs, with the coordination numbers being the odd function of the iteration order. We have found, using this procedure, that the discrete equation permits a transition to chaotic dynamics under a set of nonrestrictive initial conditions involving both the nonlinearity parameter and the width of field distribution. A peculiar feature concerning this equation is that the chaos is self-reinforcing (in view of the dependence of the Chirikov overlap parameter on the number of states) and leads to a chaotic spreading of the nonlinear field to long distances under the action of the nonlinear term. 

From a basic kinetic point of view, the spreading process is found to be non-Gaussian and non-diffusive, with complex microscopic organization revealing the presence of multiple trapping phenomena in wave number space. Most interestingly, the latter phenomena, which we have associated with the attractive interaction between unstable modes, are found to compete with some nonlocal features, consistent with L\'evy flights. 

Theoretically, the origin of those features has constituted exciting and very nontrivial a problem. It has been our proposal that the nonlocality results from insufficient embedding of the higher-order Cayley graphs into the primary order phase space, leading to the occurrence of degenerate states. In this fashion, we could explain L\'evy flights in the NLSE system~(\ref{1}) from a purely topological perspective by associating them with instantaneous jumps along degenerate states starting from the second iteration order. 

Developing these viewpoints, we have shown that a kinetic description of the asymptotic transport corresponds to a generalized diffusion equation, Eq.~(\ref{FDEL}), with fractional-derivative operators over both time (trappings, $\alpha < 1$) and the position coordinate in wave number space (flights, $\mu < 2$). In that regard, it is worthwhile to stress that NLSE~(\ref{1}) with subquadratic power nonlinearity offers an environment to generate L\'evy flights {\it dynamically} through the dependence of the Chirikov overlap parameter on the number of states (that is, without introducing the nonlocal features from the outset, conversely to the approaches of Refs. \cite{UFN,Korabel2007}). To this end, one shows based on the jump-length distribution in Eq.~(\ref{LW}) that the L\'evy index $\mu$ is directly related to the exponent of subquadratic power, $s$, via $\mu = 2s$.   

If $s\rightarrow 1$, then the multinomial expansion in Eq.~(\ref{Multinom}) is exact in the first order. That means that no degenerate states are to be expected for NLSE~(\ref{QNLSE}) [at contrast to its subquadratic generalization in Eq.~(\ref{1})]. 
As a consequence, one finds, for $s \rightarrow 1$, that L\'evy flights transform into the familiar next-neighbor random walk, leading to a pure non-Markovian dynamics with a distribution of trapping times [{\it i.e.}, Eq.~(\ref{WT}) with either $\alpha = 1/2$ or $\alpha = 1/3$]. A remarkable feature here is that quadratic nonlinearity, with $s=1$, proves to be special in that it allows for both the strong ($\alpha = 1/2$) and weak ($\alpha = 1/3$) chaos regimes depending on the strength of nonlinear interaction. This is different from the subquadratic power case, with $s < 1$, where the dependence of the nonlinear frequency shift on the width of field distribution suppresses the weak chaos regime, leading to strong chaos in the limit $t\rightarrow+\infty$. 

Concerning the onset of chaotic spreading in NLSE~(\ref{QNLSE}), we have seen that the phenomenon is critical ({\it i.e.}, thresholded), corresponding to the very special value of the nonlinearity parameter $\beta_c = 1/ \ln 2 \approx 1.442695\dots$ This value, which is a hard result of the transport model, is a characteristic of mean-field percolation on trees and pertains (strictly) to the nonlinear Schr\"odinger lattices with $s=1$ (not $s < 1$). At the same time, the dependence of the probability of site occupancy on $\beta$ in Eq.~(\ref{Exp}) indicates that the behavior is non-perturbative, as it has to be in vicinity of the criticality \cite{Sornette2004}. 

If $s=1$ exactly, then in the weak chaos regime we predict a subdiffusive spreading of the nonlinear field in accordance with the scaling law $\langle (\Delta n)^2 (t) \rangle \propto t^{1/3}$ for $t\rightarrow+\infty$. The latter scaling is different (both mathematically and dynamically) from the alternative scaling $\langle (\Delta n)^2 (t) \rangle \propto t^{1/2}$ ($t\rightarrow +\infty)$, when the chaos is strong.  No percolation-like transport has been found for $s < 1$, while in the strong chaos regime we have, for all $1/2 \leq s \leq 1$, $\langle (\Delta n)^2 (t) \rangle \propto t^{1/(s+1)}$ ($t\rightarrow+\infty$). 

Finally, we note that our results agree well with known numerical estimates, {\it e.g.}, those reported in Ref. \cite{Skokos18} for both the strong ($\alpha = 1/2$) and weak ($\alpha = 1/3$) chaos regimes. Further agreement might be concluded from the computer work of Ref. \cite{PRE21} (see Sec.~V thereof), where the plasma analogue of Anderson problem has been studied numerically on the basis of gyrokinetic approach. The results of that work suggest that systems of strongly coupled transport barriers in tokamak L-mode might decay via a subdiffusive spreading in radial direction following the scaling law in Eq.~(\ref{Sub}). In this respect, we should stress that the distinction between the chaotic and weakly chaotic dynamics remains an important and delicate problem, if only due to the borderline character of weak chaos \cite{Report,Mod} as well as the natural difficulties at studying the asymptotic transport laws.

\acknowledgments
Stimulating discussions with P. Diamond, G. Dif-Pradalier and J. Juul Rasmussen are gratefully acknowledged. AVM and AI thank the Max Planck Institute for the Physics of Complex Systems (MPI PKS, Dresden, Germany) for hospitality and financial support. 




\bibliographystyle{elsarticle-num}
\bibliography{<your-bib-database>}

\begin{thebibliography}{0}
\bibitem{Comm1}
According to the Gelfand-Naimark theorem, an arbitrary $C^*$-algebra is isomorphic to a $C^*$-subalgebra of bounded operators on a Hilbert space. In the present case, the corresponding Hilbert space is that of complex valued functions with norm and inner product.
\bibitem{Comm2}
Setting $\alpha = s/(s+1)$ to $1/3$, one gets $s=1/2$, which recovers the lower bound on $s$ in Eq.~(\ref{Iff}). Yet, the onset transport at $s=1/2$ corresponds to subquadratic nonlinearity and as such must involve L\'evy flights. This is at odds with percolation transport, considered here, which is built on random transitions between next-neighbor states and does not involve the flights.
\end{thebibliography}







\end{document}